\newcommand{\out}[1]{}
\RequirePackage{fix-cm}
\documentclass[twocolumn]{svjour3}          

\usepackage{amsfonts}
\usepackage{booktabs}
\usepackage{bussproofs}
\usepackage{float}
\usepackage{fancyvrb}
\usepackage{graphicx}
\usepackage[latin1]{inputenc}
\usepackage{listings}
\usepackage[pdftex,bookmarks=true]{hyperref}
\usepackage{pifont}
\usepackage{subfig}
\usepackage{verbatim}
\usepackage{enumitem}

 \hypersetup{
    colorlinks,%
    citecolor=black,%
    filecolor=black,%
    linkcolor=black,%
    urlcolor=black
}

\newcommand {\sq}{SPARQL}
\newtheorem{mydef}{Definition}

\newcounter{Querycount}
\newenvironment{myquery}
{
\stepcounter{Querycount} {\noindent Query} \arabic{Querycount} -
}
{
 }

\newcounter{Querysetcount}
\newenvironment{myqueryset}
{
}
{
 }

\newcounter{Datasetcount}
\newenvironment{mydataset}
{
}
{
 }

\newcommand{\tickYes}{\checkmark}
\newcommand{\tickNo}{\hspace{1pt}\ding{55}}
\begin{document}

\title{Views over RDF Datasets: A State-of-the-Art and Open Challenges}



\author{Lorena Etcheverry \and Alejandro Vaisman}


\institute{Lorena Etcheverry \at
              Instituto de Computaci\'{o}n, Facultad de Ingenier\'{i}a, Universidad de la Rep\'{u}blica\\
              \email{lorenae@fing.edu.uy}           
           \and
           Alejandro Vaisman \at
              Universit\'{e} Libre de Bruxelles and \\
              Instituto de Computaci\'{o}n, Facultad de Ingenier\'{i}a, Universidad de la Rep\'{u}blica\\
              \email{avaisman@ulb.ac.be}
}

\date{Received: date / Accepted: date}

\maketitle

\begin{abstract}
Views on RDF datasets have been discussed in several works, nevertheless there is no consensus on their definition nor the requirements they should fulfill. 
In traditional data management systems, views have proved to be useful in different application scenarios such as data integration, query answering, data security, and query modularization. 

In this work we  have reviewed existent work on views over RDF datasets, and   discussed the application of existent view definition mechanisms to four scenarios in which views have proved to be useful in traditional (relational) data management systems. To give a  framework for the  discussion we provided a definition of views over RDF datasets, an issue over which there is  no consensus so far.
We finally  chose the three proposals  closer to this definition, and analyzed them with respect to four selected goals.

\end{abstract}

\section{Introduction}
\label{intro}

With the advent of initiatives like Open Data\footnote{\url{http://www.opendefinition.org/}} and new data publication paradigms as Linked Data~\cite{Bizer2009}, the volume of data available as RDF~\cite{Klyne2004} datasets in the Semantic Web has grown dramatically. Projects such as the Linking Open Data community 
(LOD)\footnote{\url{http://www.w3.org/wiki/SweoIGTaskForces/CommunityProjects/LinkingOpenData}} encourage the publication of Open Data using the Linked Data principles which recommend using RDF as data publication format. By September 2010
 (last update of the LOD diagram), more than 200 datasets were available at the LOD site, which consisted of over 25 billion RDF triples.   This massive amount of semi-structured, interlinked and distributed data publicly at hand, faces the database community  with new challenges and opportunities:  published data need to be loaded, updated, and queried efficiently.  One question that immediately arises is: could traditional data management techniques be adapted to this new  context, and help us deal with problems such as data integration from heterogeneous and autonomous data sources, query rewriting and optimization, control access, data security, etc.? In particular, in this paper we address the issue of view definition mechanisms over RDF datasets. 
RDF datasets are formed by triples, where each triple \emph{(s,p,o)} represents that subject \emph{s} is related to object \emph{o} through the property \emph{p}. Usually, triples representing  schema and instance data coexist in RDF datasets (these are denoted TBox and ABox, respectively in Description Logics ontologies). 
A set of reserved words defined in RDF Schema (called the rdfs-vocabulary)\cite{Brickley2004} is used to define classes, properties, and to represent hierarchical relationships between them. For example, the triple \emph{(s, \texttt{rdf:type}, c)} explicitly states that \emph{s} is an instance of \emph{c} but it also implicitly states that object \emph{c} is an instance of \texttt{rdf:Class} since there exists at least one resource that is an instance of \emph{c} (see Section \ref{sec2:rdf}  for further  details on RDF). The standard query language for RDF data is SPARQL\cite{prud2008sparql}, which is based on the evaluation of graph patterns (see below for examples on SPARQL queries).   

Although view definition  mechanisms for RDF have been discussed  in the literature,  there is no consensus on what a view over RDF should be, and the requirements it should fulfill. Moreover, although we could expect  views to be  useful over the web of linked data, as they  have proved to be  in many traditional data management application scenarios (e.g.,   data integration, query answering) there is no evidence so far that this will be the case in the near future. 
In this work we discuss the usage of views in those scenarios, and  study  current RDF view definition mechanisms, with focus on key issues such as expressiveness, scalability, RDFs inference support and the integration of views into existent tools and platforms.

\subsection{Problem Statement and Motivation}
\label{example}

\begin{figure*}[!ht]
 \centering
 \includegraphics[width=\textwidth]{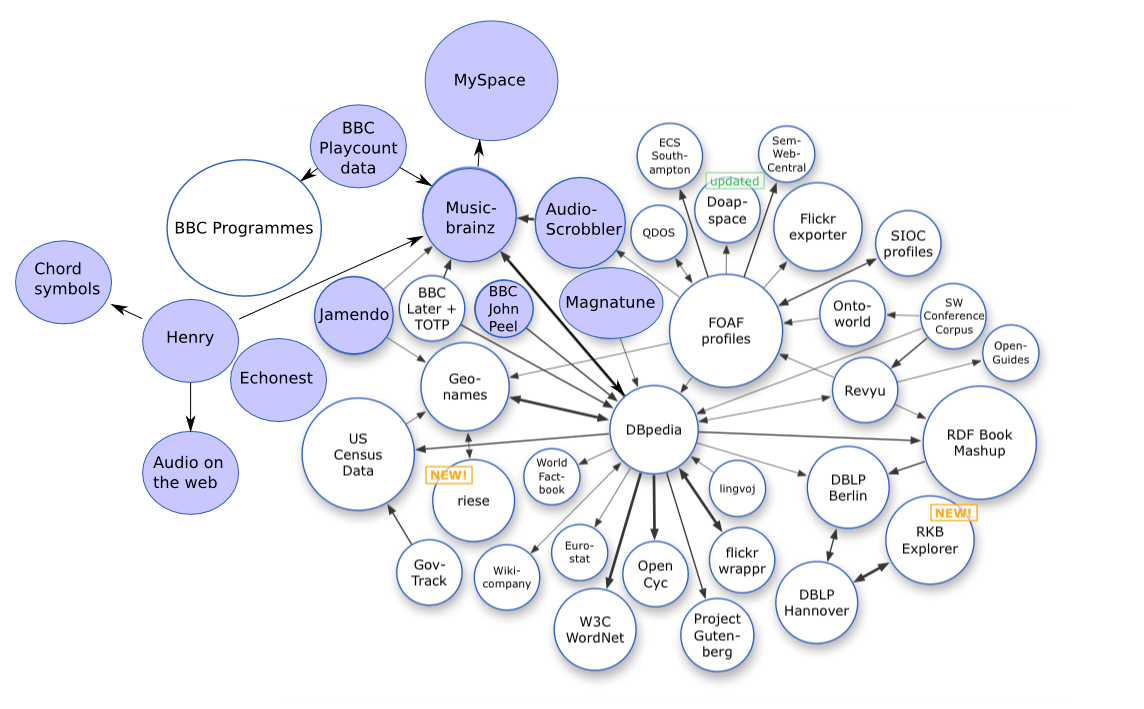}
\caption { DBTune project LOD diagram (from \url{http://dbtune.org/})}
\label{fig.lod}
\end{figure*}

The DBTune project\footnote{\url{http://dbtune.org/}} gathers more than 14 billion triples from different music-related websites. Figure~\ref{fig.lod} presents a LOD diagram that represents DBTune datasets (purple nodes), their inter-relationships and the relationships with other LOD datasets (white nodes).

\begin{figure*}
  \centering
  \subfloat[][BBC John Peel Sessions data]{\label{fig:bbc}\includegraphics[width=0.8\textwidth]{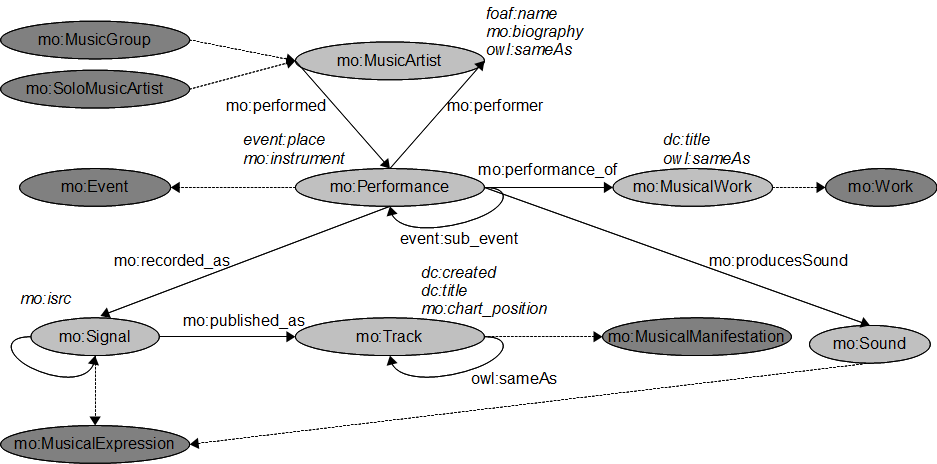}}

  \subfloat[][Jamendo website data]{\label{fig:jamendo}\includegraphics[width=0.8\textwidth]{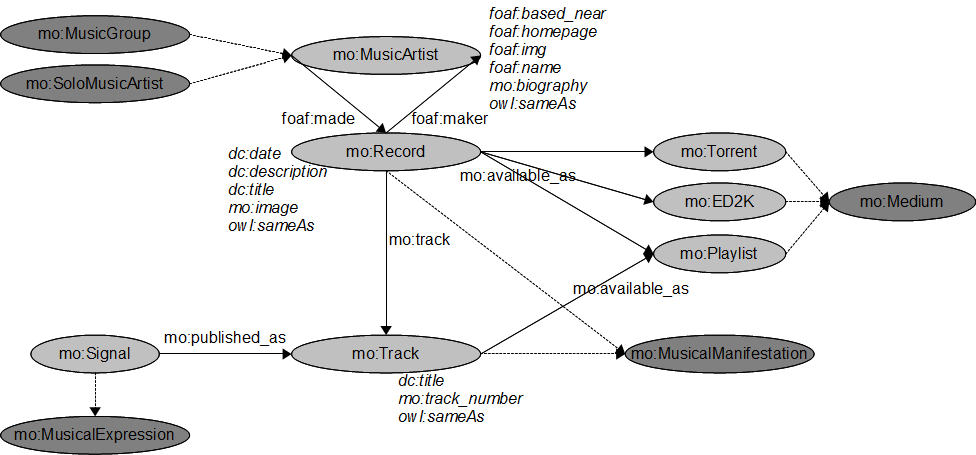}}

  \subfloat[][Magnatune record label data]{\label{fig:magnatune}\includegraphics[width=0.8\textwidth]{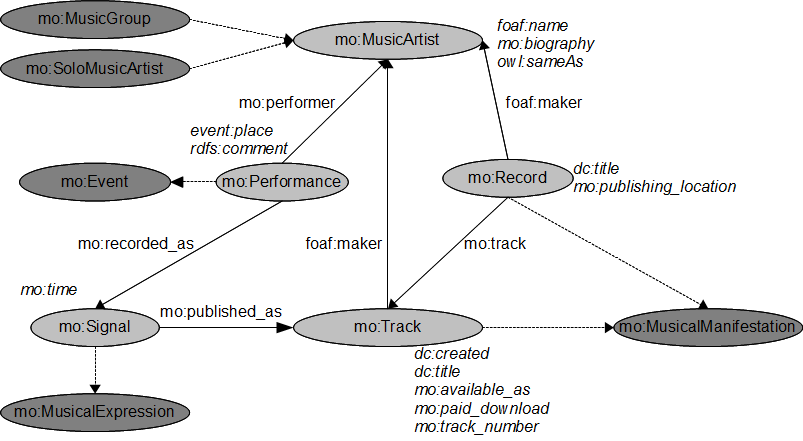}}

  \caption{Information about the schema of selected datasets from  DBTune.}
  \label{fig:sourceSch}
\end{figure*}

Each of the datasets included in the DBTune project has its own particularities. For instance,   their structures or schemas differ from each other. This is because  although DBTune  datasets are described in terms of concepts and relationships defined in the Music Ontology (MO)\footnote{\url{http://musicontology.com/}},  they do not strictly adhere to it, producing semantic and syntactic heterogeneities among them. We have selected three datasets from the DBTune project: BBC John Peel sessions dataset\footnote{\url{http://dbtune.org/bbc/peel/}}, the Jamendo website dataset\footnote{\url{http://dbtune.org/jamendo}} and the Magnatune record label dataset\footnote{\url{http://dbtune.org/magnatune}} (Section \ref{sec5:datasets} presents detailed information on this selection process, and explains the rationale behind this decision).  Information about the `schema' of the datasets can be extracted by means of   SPARQL queries. Figure \ref{fig:sourceSch} presents a graphical representation of this information. In these graphs, light grey nodes represent classes for which at least one instance is found in the dataset (we denote them used classes), dark grey nodes represent classes from the MO that are related to used classes (either as subClasses or superClasses), solid arcs represent predicates between used classes, and dashed arcs represent the
\texttt{rdfs:subClassOf} predicate. Predicates that relate classes with untyped URIs are represented in italics. 
Appendix \ref{sec:app1} describes how these graphs have been constructed.

Figure \ref{fig:sourceSch} shows that there are differences between the schemas of each data source. Let us  consider, for example, the representation of the authoring relationship between \emph{MusicArtists} and \emph{Records}. In the Jamendo dataset this relationship is represented using the \texttt{foaf:made} predicate (Figure \ref{fig:jamendo}) that connects artists with their records but also using its inverse relationship, namely the \texttt{foaf:maker} predicate between \emph{Records} and \emph{MusicArtists}. 
Although these two relationships are the inverse of each other, no assumption can be made on the consistency of data, namely that the existence of a triple (\emph{jam:artist1 foaf:made jam:record1}) does not enforce the existence of another triple of the form (\emph{jam:record1 foaf:maker jam:artist1}).
In the Magnatune dataset \emph{MusicArtists} and \emph{Records} are related using  the \texttt{foaf:maker} predicate (Figure \ref{fig:magnatune}).

We next present some use cases over the selected datasets that show how  the notion of view (in the traditional sense)   could be applied.

\textbf{Use Case 1: Retrieving artists and their records.}

A user needs  to collect information about artists and their records. To fulfill this simple requirement, a not trivial  SPARQL query must be written. This query  must  take into consideration all the different representations of the relationship between artists and records in each dataset.  Example \ref{fig.uc1select} presents a SPARQL query that returns the expected answer.

\begin{example}
A SPARQL 1.0 \texttt{SELECT} query that retrieves Artists and their Records. \\
 \lstset{language=SPARQL,framesep=4pt,basicstyle=\scriptsize,
 showstringspaces=false, tabsize=2,numbers=none,numberstyle=\small, stepnumber=1, numbersep=5pt}
  \begin{lstlisting}
SELECT DISTINCT ?artist ?record
FROM NAMED <http://dbtune.org/jamendo>
FROM NAMED <http://dbtune.org/magnatune>

WHERE { 	
	{GRAPH <http://dbtune.org/jamendo/>
		{  ?artist foaf:made ?record . 
		   ?artist rdf:type mo:MusicArtist .	
		   ?record rdf:type mo:Record }
	}UNION
	{GRAPH <http://dbtune.org/jamendo/>
		{ ?record foaf:maker ?artist .
		  ?artist rdf:type mo:MusicArtist .
		  ?record rdf:type mo:Record }
	}UNION
	{GRAPH <http://dbtune.org/magnatune/>
		{ ?record foaf:maker ?artist .
		  ?artist rdf:type mo:MusicArtist .
		  ?record rdf:type mo:Record  }
}}

  \end{lstlisting} \qed

\label{fig.uc1select}
\end{example}

SPARQL queries are too complex to be written by an end user, and require a precise knowledge of the schema.  Therefore,  it would be desirable to somehow provide a uniform representation of this relationship in order to simplify querying the integrated information.
Several strategies could be used  to provide a uniform view of all datasets. One possibility would be to materialize the missing triples, which in this case leads to the creation of new triples in the Magnatune  and Jamendo datasets. For each \emph{(record, \texttt{foaf:maker}, artist)} triple that relates a Record \emph{record} with an Artist \emph{artist}, a new triple \emph{(artist, \texttt{foaf:made}, record)} must be added to the dataset. This strategy would be hard to maintain and could also interfere with the independence of the sources.

To avoid maintenance issues, approaches that dynamically generate virtual triples are needed. Some  of them use reasoning and rules to create mappings between concepts and infer knowledge that is not explicitly stated \cite{Hogan2011}.
Another approach could be to build new graphs that encapsulate underlying heterogeneities. For instance, SPARQL \texttt{CONSTRUCT} queries return graphs dynamically created from existent ones and allow the creation of new triples as the next example shows. 

\begin{example}
The following  SPARQL \texttt{CONSTRUCT} query returns a graph that contains all the \emph{(artist, \texttt{foaf:made}, record)} triples from the Jamendo dataset but also generates new triples. That is, for each \emph{(record, \texttt{foaf:maker}, artist)} triple in the Magnatune and Jamendo datasets it creates a \emph{(artist, \texttt{foaf:made}, record)} triple) (i.e, the query of Example~\ref{fig.uc1select}).

 \lstset{language=SPARQL,framesep=4pt,basicstyle=\scriptsize,
 showstringspaces=false, tabsize=2,numbers=none,numberstyle=\small, stepnumber=1, numbersep=5pt}
 \begin{lstlisting}
CONSTRUCT {?artist foaf:made ?record}
FROM NAMED <http://dbtune.org/jamendo>
FROM NAMED <http://dbtune.org/magnatune>
WHERE { 	
	{GRAPH <http://dbtune.org/jamendo>
		{ ?artist foaf:made ?record .
		  ?artist rdf:type mo:MusicArtist .	
		  ?record rdf:type mo:Record  }
	}UNION
	{GRAPH <http://dbtune.org/jamendo>
		{ ?record foaf:maker ?artist .
		  ?artist rdf:type mo:MusicArtist .
		  ?record rdf:type mo:Record }
	}UNION
	{GRAPH <http://dbtune.org/magnatune>
		{ ?record foaf:maker ?artist .
		  ?artist rdf:type mo:MusicArtist .
		  ?record rdf:type mo:Record  }
	 }
}
 \end{lstlisting} \qed

\label{fig.uc1construct}
\end{example}

Let us suppose that now  our user wants to reutilize  this query to retrieve the title of  each record made by an artist. 
Although the query  in Example \ref{fig.uc1construct} generates a new graph,  SPARQL does not provide mechanisms to pose queries against dynamically generated 
graphs (e.g., using graphs as sub-queries in the \texttt{FROM} clause). To answer this query in SPARQL 1.0 existent 
queries cannot be reused, and a new query must be formulated (see next example).

\begin{example}
 The  SPARQL 1.0 \texttt{SELECT} query below,  retrieves artists, records and record titles.

 \lstset{language=SPARQL,framesep=4pt,basicstyle=\scriptsize,
 showstringspaces=false, tabsize=2,numbers=none,numberstyle=\small, stepnumber=1, numbersep=5pt}
\begin{lstlisting}
SELECT DISTINCT ?artist ?record ?title
FROM <http://dbtune.org/jamendo>
FROM <http://dbtune.org/magnatune>
FROM NAMED <http://dbtune.org/jamendo>
FROM NAMED <http://dbtune.org/magnatune>
WHERE { ?record dc:title ?title .
	{GRAPH <http://dbtune.org/jamendo/>
		{  ?artist foaf:made ?record . 
		   ?artist rdf:type mo:MusicArtist .	
		   ?record rdf:type mo:Record .
		}
	}UNION
	{GRAPH <http://dbtune.org/jamendo/>
		{ ?record foaf:maker ?artist .
		  ?artist rdf:type mo:MusicArtist .
		  ?record rdf:type mo:Record .
		}
	}UNION
	{GRAPH <http://dbtune.org/magnatune/>
		{ ?record foaf:maker ?artist .
		  ?artist rdf:type mo:MusicArtist .
		  ?record rdf:type mo:Record .
		}
}}
\end{lstlisting} \qed

\label{fig.uc1sparql10app}
\end{example}

The SPARQL 1.1 proposal~\cite{Harris2010}  (see Section \ref{sec2})  partially supports sub-queries, allowing only \texttt{SELECT}  queries to be part of the \texttt{WHERE} clause. Existent \texttt{CONSTRUCT} queries cannot be reused either in the \texttt{FROM} clause (e.g.: as datasets) nor in the \texttt{WHERE} clause (e.g.: as graph patterns). 
Example \ref{fig.uc1sparql11app} presents a SPARQL 1.1 \texttt{SELECT} query that retrieves artists, their records and their titles. It shows that, in order to reuse the query presented in Example \ref{fig.uc1select}, the code must be `copy-pasted', which is  hard to maintain, error-prone,
and limits the use of optimization strategies based on view materialization.

\begin{example}
A SPARQL 1.1 \texttt{SELECT} query that retrieves artists, records and record titles.
 \lstset{language=SPARQL,framesep=4pt,basicstyle=\scriptsize,
 showstringspaces=false, tabsize=2,numbers=none,numberstyle=\small, stepnumber=1, numbersep=5pt}
\begin{lstlisting}

SELECT ?artist ?record ?recordTitle
WHERE { ?record dc:title ?recordTitle .
{SELECT ?artist ?record
 FROM <http://dbtune.org/magnatune>
 WHERE { ?record foaf:maker ?artist .
         ?artist a mo:MusicArtist .
         ?record a mo:Record }
}UNION
{SELECT ?artist ?record
 FROM <http://dbtune.org/jamendo>
 WHERE { ?artist foaf:made ?record .
         ?artist a mo:MusicArtist .
         ?record a mo:Record }
}UNION
{SELECT ?artist ?record
 FROM <http://dbtune.org/jamendo>
 WHERE { ?record foaf:maker ?artist .
         ?artist a mo:MusicArtist .
         ?record a mo:Record }
}}
\end{lstlisting} \qed
\label{fig.uc1sparql11app}
\end{example}

In light of the above, SPARQL extensions have been proposed to allow \texttt{CONSTRUCT} queries to be used as subqueries. For instance, Networked Graphs (NG)~\cite{Schenk2008}  allow defining and storing graphs for later use in other queries. Example \ref{fig.uc1NGdef} shows, using RDF TriG syntax\footnote{\url{http://www4.wiwiss.fu-berlin.de/bizer/TriG/}}, how the graph in Example \ref{fig.uc1select} can be implemented using NGs. An NG is defined by means of  an RDF triple whose subject is the URI that  identifies the graph, its predicate is denoted \texttt{ng:definedBy}, and its object is a string that represents the \texttt{CONSTRUCT} query that will be evaluated
at runtime, and whose results will populate the graph.

\begin{example}
Applying Networked Graphs to Use Case 1: definition
 \lstset{language=SPARQL,framesep=4pt,basicstyle=\scriptsize,
 showstringspaces=false, tabsize=2,numbers=none,numberstyle=\small, stepnumber=1, numbersep=5pt}
\begin{lstlisting}
def:query1 {
def:query1 ng:definedBy
``CONSTRUCT {?artist foaf:made ?record}
WHERE { 	
	{GRAPH <http://dbtune.org/jamendo/>
		{ ?artist foaf:made ?record .
		  ?artist rdf:type mo:MusicArtist .	
		  ?record rdf:type mo:Record
		}
	}UNION
	{GRAPH <http://dbtune.org/jamendo/>
		{ ?record foaf:maker ?artist .
		  ?artist rdf:type mo:MusicArtist .
		  ?record rdf:type mo:Record
		}
	}UNION
	{GRAPH <http://dbtune.org/magnatune/>
		{ ?record foaf:maker ?artist .
		  ?artist rdf:type mo:MusicArtist .
		  ?record rdf:type mo:Record 
		}
	}}''^^ng:query
}

\end{lstlisting} \qed

\label{fig.uc1NGdef}
\end{example}

Once defined,  the NG can be reused in further queries. Example \ref{fig.uc1NGapp} presents a SPARQL query that uses the previously defined NG, encapsulating the different representations of the relationship between artists and their records.

\begin{example}
Applying Networked Graphs to Use Case 1: usage
 \lstset{language=SPARQL,framesep=4pt,basicstyle=\scriptsize,
 showstringspaces=false, tabsize=2,numbers=none,numberstyle=\small, stepnumber=1, numbersep=5pt}
\begin{lstlisting}

SELECT DISTINCT ?artist ?record ?recordTitle
WHERE	{ ?record dc:title ?recordTitle .
		{ GRAPH <http://definedViews/query1>
			{?artist foaf:made ?record }        
   		}
}
\end{lstlisting}\qed
\label{fig.uc1NGapp}
\end{example}

\textbf{Use Case 2: Musical manifestations and their authors.}

Let us now consider that the user wants to retrieve information about  all musical manifestations stored in the datasets. 
Figure~\ref{fig:sourceSch} shows that there are no instances of the \emph{MusicalManifestation}  
class in the datasets but there are instances of two of their sub-classes: \emph{Record} and \emph{Track}.
SPARQL  supports different entailment regimes, in particular RDF, RDFS, and OWL\footnote{\url{http://www.w3.org/TR/owl-features/}}. Under RDFS entailment the application of inference rules generates results that are not explicitly stated in the datasets. 
For example,  one of such rules  allows inferring that, since \emph{Record} and \emph{Track} are sub-classes of \emph{MusicalManifestation}  all the instances of \emph{Record} and \emph{Track} are also instances of \emph{MusicalManifestation}. We take a closer look at inference mechanisms in Section \ref{sec2:rdf}

Example \ref{fig.usecase2} shows a SPARQL \texttt{CONSTRUCT} query that creates a graph that contains all the Musical Manifestation instances and for each instance its author, in case  available. Since  \emph{Record} and \emph{Track} are sub-classes of \emph{MusicalManifestation}, all instances of the former two are also instances of the latter. Thus, they should appear in the resulting graph. This query can be stored using NGs or implemented using SPARQL++ ~\cite{Polleres2007}. We discuss SPARQL++ later in this paper.

\begin{example}
 Musical manifestations and their authors.
 \lstset{language=SPARQL,framesep=4pt,basicstyle=\scriptsize,
 showstringspaces=false, tabsize=2,numbers=none,numberstyle=\small, stepnumber=1, numbersep=5pt}
 \begin{lstlisting}
 CONSTRUCT {	
	?mm rdf:type mo:MusicalManifestation .
	?mm foaf:maker ?artist }
WHERE {	?mm rdf:type mo:MusicalManifestation .
		OPTIONAL{
			?mm foaf:maker ?artist } .
		OPTIONAL{
			?mm a mo:Track .
			?record mo:track ?mmanifestation .
			?record foaf:maker ?artist } .               
        }
 \end{lstlisting}\qed
 \label{fig.usecase2}
\end{example}

This use case exemplifies a problem orthogonal to the one stated in Use Case 1: the need of support entailment regimes  in SPARQL implementations and  in view definition mechanisms.
Although these mechanisms, at first sight, seem to solve the problems above, little information can be found in the literature regarding how to use them, the volume of data they can handle and also on the restrictions that may apply to the queries they support.

The purpose of this work is two-fold. First, study different application scenarios in which views over RDF datasets could be useful;  second, discuss to what extent existent view definition mechanisms can be used on the described scenarios.

\subsection{Contributions and Paper Organization}
\label{sec1:contribs}

 This paper is aimed at  providing an  analysis  of the state-of-the-art in view definition mechanisms over RDF datasets,
 and identifying open research problems in the field. 
 We first introduce the  basic concepts on RDF, RDFS and SPARQL   (Section \ref{sec2}). 
 In Section~\ref{sec3},  to give 
 a framework to our study, we propose a definition of views over RDF datasets, along with four  scenarios in which 
 views have been  traditionally applied in relational database systems.    
 In Section \ref{sec4}  we study 
 current view definition mechanisms, with a focus on the three ones that fulfill most of the conditions of our 
 definition of views, and support the scenarios mentioned above. These proposals are 
 SPARQL++, Networked  Graphs, and vSPARQL. We  also provide a wider view, discussing 
 other proposals in the field. In Section \ref{sec5} we analyze the three selected proposals with respect to four goals: SPARQL 1.0  
  support, inference support, scalability, and facility for integration with existent platforms. We also perform experiments over 
  the current  a Networked Graphs implementation. 
  Finally, in Section \ref{sec6} we present our conclusions and analyze open research  directions.

\section{Preliminaries}
\label{sec2}

To make this paper self-contained in this section we present a brief review of basic concepts on RDF, RDFS and SPARQL~\cite{Angles2008,Arenas2009,Glimm2011,Hitzler2009}.

\subsection{RDF and RDFS}
\label{sec2:rdf}

The Resource Description Framework (RDF)~\cite{Klyne2004} is a data model for expressing assertions over resources identified by an universal resource identifier (URI). Assertions are expressed as \emph{subject-predicate-object} triples, where \emph{subject} are always resources, and \emph{predicate} and \emph{object} could be resources or strings. \emph{Blank nodes} (\emph{bnodes}) are used to represent anonymous resources or resources without an URI, typically  with  a structural function,  e.g., to group a set of statements. Data values in RDF are called \emph{literals} and can only be \emph{objects} in triples. A set of RDF triples or \emph{RDF dataset} can be seen as a directed graph where \emph{subject} and \emph{object} are nodes, and \emph{predicates} are arcs. Formally:

\begin{mydef}[RDF Graphs]
Consider the following sets U (URI references); 
$ B= \lbrace N_{j} \in \mathbb{N} \rbrace $
(blank nodes); and L (RDF literals). A triple $(v1, v2, v3) \in (U\cup B) \times U \times (U \cup B \cup L)$ is called an RDF triple. We denote UBL the   union  U $\cup$ B $\cup$ L. An RDF graph is a set of RDF triples. A subgraph is a subset of a graph. A graph is
 \emph{ground} if it has no blank nodes. \qed
\end{mydef}

Although the standard RDF serialization format is RDF/XML~\cite{Beckett2004}, several formats coexist in the web such as NTriples\cite{Beckett2004a}, Turtle~\cite{Beckett2011}, N3~\cite{Berners-Lee2006}, Trig~\cite{Bizer2007}, and several serialization formats over JSON~\cite{W3C2010}.

RDF Schema (RDFS)~\cite{Brickley2004} is a particular RDF vocabulary  supporting inheritance of classes and properties, as well as typing, among other features. In this work we restrict ourselves  to a fragment of this vocabulary which includes the most used features of RDF, contains the essential semantics, and is computationally more efficient than the complete RDFS vocabulary~\cite{Munoz2009}
This fragment, called $\rho$df, contains the following predicates: rdfs:range \texttt{[range]}, rdfs:domain \texttt{[dom]}, rdf:type \texttt{[type]}, rdfs:subClassOf \texttt{[sc]}, and rdfs:subPropertyOf \texttt{[sp]}.  The following set of rules captures the semantics of $\rho$df and allows reasoning over RDF. Capital letters represent variables to be instantiated by elements of UBL. 
 We use  this subset of RDFS for addressing inference capabilities in view definitions.

\textbf{Group A (Subproperty)}
\begin{prooftree}
\AxiomC{(A, \texttt{sp}, B)}
\AxiomC{(B, \texttt{sp}, C)}
\RightLabel{(1)}
\BinaryInfC{(A,\texttt{sp}, C)}
\end{prooftree}

\begin{prooftree}
\AxiomC{(A, \texttt{sp}, B)}
\AxiomC{(X, A, Y)}
\RightLabel{(2)}
\BinaryInfC{(X, B, Y)}
\end{prooftree}

\textbf{Group B (Subclass)}
\begin{prooftree}
\AxiomC{(A, \texttt{sc}, B)}
\AxiomC{(B, \texttt{sc}, C)}
\RightLabel{(3)}
\BinaryInfC{(A, \texttt{sc}, C)}
\end{prooftree}

\begin{prooftree}
\AxiomC{(A, \texttt{sc}, B)}
\AxiomC{(X, \texttt{type}, A)}
\RightLabel{(4)}
\BinaryInfC{(X, \texttt{type}, B)}
\end{prooftree}

\textbf{Group C (Typing)}
\begin{prooftree}
\AxiomC{(A, \texttt{dom}, C)}
\AxiomC{(X, A, Y)}
\RightLabel{(5)}
\BinaryInfC{(X, \texttt{type}, C)}
\end{prooftree}

\begin{prooftree}
\AxiomC{(A, \texttt{range}, D)}
\AxiomC{(X, A, Y)}
\RightLabel{(6)}
\BinaryInfC{(Y, \texttt{type}, D )}
\end{prooftree}

\subsection{SPARQL}
\label{sec2:sparql}

SPARQL is a query language for RDF graphs, which became a W3C standard  in 2008~\cite{prud2008sparql}. The query evaluation  mechanism  of  SPARQL is based on subgraph matching: RDF triples in the queried data and a query pattern are interpreted as nodes and edges of directed graphs, and the   query graph is matched  to the data graph, instantiating  the variables in the query graph definition~\cite{Glimm2011}. The selection criteria is expressed as a graph pattern in the \texttt{WHERE} clause, and it is composed of basic graph patterns  defined as follows:
 
 \begin{mydef}[Queries]
\sq\ queries are built using an infinite set V of  variables disjoint from UBL. A variable  v $\in$ V is denoted using either ? or \$ as a prefix. A triple pattern is member of the set $(UBL \cup V) \times (U \cup V) \times (UBL \cup V)$, that binds variables in V to RDF Terms in the graph. A basic graph pattern (BGP) is a set of triple patterns connected by the `.' operator. \qed
\end{mydef}

Complex graph patterns can be built starting from BGPs, which include:
\begin{itemize}
\item \textbf{group graph patterns}, a graph pattern containing multiple graph patterns that must all match,
\item \textbf{optional graph patterns}, a graph pattern that may match and extend the solution, but will not cause the query to fail,
\item \textbf{union graph patterns}, a set of graph patterns that are tried to match independently, and
\item \textbf{patterns on named graphs}, a graph pattern that is matched against named graphs.
\end{itemize}

SPARQL queries have four query forms. These query forms use  variable bindings to create the results of the query. The query forms are: 
\begin{itemize}
\item \texttt{SELECT}, which returns a set of the variables bound in the query pattern, 
\item \texttt{CONSTRUCT}, which returns an RDF graph constructed by substituting variables in a set of triple templates,
\item \texttt{ASK} , which returns a boolean value indicating whether a query pattern matches or not, and
\item \texttt{DESCRIBE}, which returns an RDF graph that describes resources found.
\end{itemize}

Table \ref{tab.sparqlStr} presents a summary of the structure of queries in SPARQL 1.0\footnote{Adapted from \url{www.dajobe.org/2005/04-sparql/SPARQLreference-1.8.pdf}} where every part of the query is optional, except for the results format clause.  

\begin{table}[htbp]
\caption{SPARQL 1.0 query structure}
\label{tab.sparqlStr}
\begin{tabular}{|p{1.5cm}|p{6cm}|}
\hline
Prologue & BASE \textless URI\textgreater \\ \cline{ 2- 2}
{} & PREFIX prefix: \textless URI\textgreater (repeatable) \\ \hline
Result format (required) & SELECT (DISTINCT) [sequence of ?variable $\vert$ *] \\ \cline{ 2- 2}
{} & DESCRIBE [sequence of ?variable $\vert$ * $\vert$ \textless URI\textgreater] \\ \cline{ 2- 2}
{} & CONSTRUCT \{ graph pattern \} \\ \cline{ 2- 2}
{} & ASK \\ \hline
Dataset Sources & FROM \textless URI\textgreater (Adds triples to the background graph, repeatable) \\ \cline{ 2- 2}
{} & FROM NAMED \textless URI\textgreater (Adds a named graph, repeatable) \\ \hline
Graph Pattern & WHERE \{ graph pattern [ FILTER expression ]\} \\ \hline
Results Ordering & ORDER BY sequence of ?variable \\ \hline
Results Selection & LIMIT n, OFFSET m \\ \hline
\end{tabular}
\end{table}

\subsection{SPARQL 1.1~~}
The SPARQL 1.1 specification~\cite{Harris2010},  with status of working draft at the moment of writing this paper, 
 includes several functionalities that extend the query language power. We next summarize the most relevant ones.
 
\begin{itemize}
\item Sub-queries: more specifically sub-select queries in the \texttt{FROM} clause;
\item Aggregates: \texttt{GROUP BY} clause and aggregate expressions in \texttt{SELECT} clause, such as \texttt{AVG}, \texttt{COUNT}, \texttt{MAX}, etc.;
\item  New mechanisms for negation and filtering besides traditional negation by failure (already available in SPARQL 1.0), 
e.g.,   \texttt{NOT EXISTS} expressions within \texttt{WHERE} clauses are introduced;
\item Property paths: SPARQL 1.1  allows property paths, which specify a possible route between nodes in a graph. 
Property paths are similar to XPath expression in XML.
\item Variables:  new variables may be introduced within queries or results, e.g.: \texttt{SELECT (expr AS ?var)} allows  projecting
 a new variable into the result set, while \texttt{BIND (expr AS ?var)} can be used to assign values to variables, 
\end{itemize}

\section{RDF Views: Definition and Scenarios}
\label{sec3}

Views over RDF datasets have been discussed in several works,  although  there is not yet a consensus about their  definition and characterization.
Some of these works are  not based on SPARQL, but   provide useful insight on the problem at hand. In particular, in~\cite{Magkanaraki2004} the authors propose RVL, a view definition language based on RQL query language. RVL views  
enforce the separation between schema and data, specifying a virtual schema with new RDFS classes and properties and a set of graph patterns that allow the computation of instances. RVL view definitions can be stored and used in other queries.
In~\cite{Volz2003} the authors claim that, from the perspective of classical databases, views can be considered as arbitrary stored queries, but no conceptual description of views is provided. On the other hand, they state that views in the Semantic Web must have a precise semantics described by an ontology, which should also embed the view in its appropriate location within the inheritance hierarchies.

Recent work based on SPARQL lacks of a clear definition of views ~\cite{Polleres2007,Schenk2008,Shaw2011}. Even some of these proposals actually extend SPARQL query capabilities, not giving an adequate argumentation about why those new features are required in a view definition language. 

In our approach, an RDF view must meet the following requirements:
\begin{enumerate}
\item Should be specified using SPARQL;
\item The result of the evaluation of an RDF view over an RDF graph should be an RDF graph, obtained using SPARQL semantics;
\item The result of the evaluation of an RDF view  should consider RDF and RDFS entailment regimes;
\item It should be possible to store RDF views for later use as sub-queries;
\end{enumerate}

According to these requirements, we provide the following definition:

\begin{mydef}[RDF Views]
\label{def:views}
An RDF  view V  is a pair V = (n; $Q_v$), where n is a URI denoting the name of the view, and $Q_v$ is a SPARQL \texttt{CONSTRUCT} query that defines the structure and the contents of the view V. \qed
\end{mydef}


\subsection{Application Scenarios}
\label{sec3:scenarios}

Although Semantic Web based data management systems seem to pose new problems and challenges to the research community, we believe that some ideas can be brought from traditional database systems to solve known problems in this new context.
In particular, views, and more specifically relational database views, play an important role in different application scenarios in traditional data management systems. Within relational databases, view definition languages make it possible to select and (with some limitations) modify the data needed by an application without materializing it; then queries are written using the defined view and evaluated against the original dataset. View specification in SQL allows  defining the schema of the view and the instances that will populate it, based not only on the underlying schema and its instances but also allowing the creation of new columns and instances, using built-in transformation functions (e.g., concatenate) or aggregate functions. As stated in ~\cite{Halevy2001} much of the work on relational views has focused on Select-Project-Join (SPJ) queries, but numerous extensions have been proposed for queries including grouping, aggregation and multiple SQL blocks, recursive queries, views with access-pattern limitations, queries over object-oriented databases and queries over semi-structured data. We  now define four classic application scenarios where views have been proved useful in relational databases, analyze  those scenarios  in the Semantic Web context, and  study how views characterized by Definition \ref{def:views} can be applied to    them.  In Section \ref{sec4} we study how existing proposals  are suitable to solve the problems that arise in these described scenarios.

\paragraph*{Scenario 1: Views and data integration.~~}
Traditionally, data integration systems make extensive use of views to  provide a reconciled and integrated vision of the underlying data sources. Well-known approaches, based on virtual data  integration, use the idea of creating a global or mediated schema and either expressing local data sources as views over the mediated schema (Local As View LAV), expressing global schema as views over local data sources (Global As View GAV) or hybrid approaches such as GLAV~\cite{Lenzerini2002}. Data warehouses and federated database systems are examples of traditional data integration systems.
Schema matching and resolving mappings between the global schema and the sources are key issues in this scenario. Dealing with inconsistencies between sources, semantic heterogeneity and query optimization are also interesting problems in data integration systems.

Semantic web data integration is an active area of research that faces important challenges and also presents several research opportunities as data on the web is inherently heterogeneous, either semantically and syntactically, messy, inconsistent, volatile and big. At least three different approaches can be distinguished: virtual integration, materialized integration and hybrid.
Within the  virtual integration approach the idea is the same as in traditional data management systems: to transform the source datasets to a common schema or representation without materializing those triples. Networked Graphs~\cite{Schenk2008}  (which we comment below) allows performing this kind of virtual data integration and use case 1 presented in Section~\ref{example} is an example of its application to a simple data integration task.
The approach presented~\cite{Hogan2011} can be seen as an hybrid one, since on the one hand transformation or views are specified using rules but inferred triples are materialized for later user. By doing time-consuming reasoning tasks off-line the authors improve the performance,  one of the big issues related to web-scale reasoning techniques.
Some authors argue on the applicability of traditional data integration techniques to this context. For instance, Dataspace Support Platforms (DSSP) propose an evolving data integration system which tries to distribute over time the modeling costs inherent to data integration problems in a pay-as-you-go fashion~\cite{Sarma2011}.

\paragraph*{Scenario 2: Query answering using views.~~}
Materialized views or indexes can be used to optimize query computation. For this, queries must be completely or partially rewritten in terms of existent views. In traditional data management systems the problem of finding re-writings, highly related to the problem of query containment, has been widely studied. In~\cite{Halevy2001,Levy1995}  the authors define the problems of finding a rewriting of a query in terms of views, finding a minimal rewriting and completely resolving a query using views, also analyzing the complexity of those problems. They prove that the problem of finding a minimal rewriting for conjunctive queries with no built-in predicates is NP-complete.

The problem of query answering using views has also been translated into the Semantic Web context. This problem can be decomposed in two sub-problems: centralized query answering and distributed query answering. With respect to the former, 
several works support    query answering using views in a centralized context through the notion of indexing~\cite{Castillo2010,Dritsou2011,Neumann2010}. We comment on them in Section \ref{sec4:index}.   
On the other hand, current implementations of Semantic Web search engines,  tend to reduce the problem of distributed query answering to centralized query answering. They apply ideas from relational data warehouses and search engines, crawling RDF datasets for materialization and indexing in a centralized data store~\cite{Cheng2009,Aquin2007,Lopez2010,Oren2008a}.
In \cite{Harth2010a} the authors propose an hybrid approach. They designed a mechanism to perform the selection of relevant sources for a certain query in distributed query processing. They build and maintain data summaries for each source, which are used in the selection process, and then retrieve the RDF data from the selected sources into main memory in order to perform join operations.

Regarding SPARQL query optimization a thoughtful analysis of complexity and strategies has been made in~\cite{Schmidt2010}.

\paragraph*{Scenario 3: Views and data security.~~}
In traditional data management systems views have been used to implement security policies and restrictions over data access~\cite{Denning1986}.
Also in the context of XML data, views as XPath queries have been used to implement control access policies~\cite{Damiani2002}
 
 A direct application of views to this problem can be found in ~\cite{Flouris2010}, where the authors present an access control specification language that allows to define triple-level authorisation permissions. Their work is based on the specification of control access permissions as sets of triples that satisfy certain graph pattern. These sets are either annotated as included or excluded from result sets. Control access permissions are implemented as named graphs and queries are performed over them.
Several works can be found in the literature regarding RDF data access control policies and trust management~\cite{Abel2007,Artz2007,Gil2011,Golbeck2003}.

\paragraph*{Scenario 4: Views and query modularization.~~}
 Views and subqueries are also used to make complex queries easier to understand. However,  these improvements in readability may lead to downgrades in performance if rewriting tasks are not performed adequately.

In the case of SPARQL queries, the ability to include queries in the \texttt{FROM} and \texttt{FROM NAMED} clauses leads to query composition and modularization, also allowing the optimization of queries since selection and projection can be pushed down in the evaluation tree of a query~\cite{Angles2011}. The next example illustrates this issue. 

\begin{example} [Query modularization] 
\label{ex:example8}
The following   query retrieves pairs of names of artists who have performed in the same location. The inner \texttt{CONSTRUCT} query returns a graph with pairs of artist that have performed in the same location.

 \lstset{language=SPARQL,framesep=4pt,basicstyle=\scriptsize,
 showstringspaces=false, tabsize=1,numbers=none,numberstyle=\small, stepnumber=1, numbersep=5pt}
 
 \begin{lstlisting}
SELECT ?name1 ?name2
FROM dbtune:peel
FROM
	( CONSTRUCT {?a1 def:coleagues ?a2}
	  WHERE {?a1 mo:performed ?p1 .
	   ?a2 mo:performed ?p2 .
	   ?p1 event:place ?pl1 .
	   ?p2 event:place ?pl1 .
	   FILTER(!(?a1 = ?a2)) }
	)
WHERE {?artist1 def:coleagues ?artist2 .
       ?artist1 foaf:name ?name1 .
       ?artist2 foaf:name ?name2 
       }

 \end{lstlisting}\qed
\label{fig:qmodul1}
\end{example}

Provided that  the query language supports  it, the inner query could be replaced by a view that could either be executed at runtime, or by a 
materialized view. We study languages supporting this feature in Section \ref{sec4}.

\section{Existing Proposals for  RDF Views}
\label{sec4}

In the following we discuss different approaches related to the notion of view, 
and how they address the scenarios defined in Section \ref{sec3}.
From these approaches, we then select and discuss the ones that are closest to our vision of what a view in RDF should be (Definiton \ref{def:views}), namely   SPARQL++~\cite{Polleres2007}, Networked Graphs (NG)~\cite{Schenk2008}, and vSPARQL~\cite{Shaw2011}. 
Other proposals, not so closely related to our definition,  are also briefly commented.

\subsection{SPARQL++}
\label{sec4:sparql++}

Polleres et al.~\cite{Polleres2007} propose extensions to SPARQL 1.0 that not only include the capability of using nested \texttt{CONSTRUCT} queries in \texttt{FROM} clauses but also allow to define built-in and aggregation functions and function calls in the \texttt{CONSTRUCT} clause.
The implementation is based on the translation of SPARQL++ queries into HEX-programs, an extension of logic programs under answer-set semantics~\cite{Polleres2007a}. The translated queries are then processed using dlvhex, an HEX-program solver based on DLV\footnote{\url{http://www.dbai.tuwien.ac.at/proj/dlv/}} which is a disjunctive Datalog system. The source code is available online\footnote{\url{http://sourceforge.net/projects/dlvhex-semweb/}}.

SPARQL++ queries cannot be stored for use in other queries. In order to reuse the queries must be `copy-pasted'.
With  respect to the scenarios defined in Section~\ref{sec3}, and due to its inability to store views definitions, SPARQL++ partially supports scenarios 1 (view integration), and 4 (query modularization). Moreover, the query in Example \ref{ex:example8}  is compliant with SPARQL++ syntax. In the following example we show how to apply SPARQL++ to use case 1.

\begin{example}[Applying SPARQL++ to Use Case 1]

 \lstset{language=SPARQL,framesep=4pt,basicstyle=\scriptsize,
 showstringspaces=false, tabsize=2,numbers=none,numberstyle=\small, stepnumber=1, numbersep=5pt}
 \begin{lstlisting}
SELECT DISTINCT ?artist ?record ?recordTitle
WHERE {
   ?record dc:title ?recordTitle .
   {CONSTRUCT {?artist foaf:made ?record}
   WHERE { 	
	{GRAPH <http://dbtune.org/jamendo/>
		{ ?artist foaf:made ?record .
		  ?artist rdf:type mo:MusicArtist .	
		  ?record rdf:type mo:Record
		}
	}UNION
	{GRAPH <http://dbtune.org/jamendo/>
		{ ?record foaf:maker ?artist .
		  ?artist rdf:type mo:MusicArtist .
		  ?record rdf:type mo:Record
		}
	}UNION
    {GRAPH <http://dbtune.org/magnatune/>
      { ?record foaf:maker ?artist .
        ?artist rdf:type mo:MusicArtist .
        ?record rdf:type mo:Record 
		}
	}
}}}
 \end{lstlisting}\qed
\end{example}

\subsection{Networked Graphs}
\label{sec4:NG}

Schenck et al.~\cite{Schenk2008} propose Networked Graphs, a declarative mechanism to define RDF graphs as \texttt{CONSTRUCT} queries and \emph{named graphs}. Networked Graphs (NG) support negation, as available in SPARQL 1.0 (negation by failure) and also queries that use NGs distributed over different endpoints. The semantics of NGs is an adaptation of the well founded semantics (WFS) for logic programs and the algorithm that performs the evaluation uses a variation of the alternating fixpoint algorithm for computing WFS. NGs  implementation supports cycles.

An NG is defined by means of an RDF triple whose subject is the URI that identifies the graph, its predicate is denoted \texttt{ng:definedBy}, and its object is a string that represents the \texttt{CONSTRUCT} query that will be evaluated on runtime, and whose results will populate the graph.

\begin{figure}[!ht]
 \centering
 \includegraphics[scale=0.7]{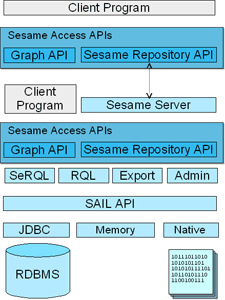}
\caption { Sesame architecture (from \url{http://www.openrdf.org/doc/sesame/users/userguide.html/})}
\label{fig:sesame}
\end{figure}

The implementation of NGs is based on Sesame 2 RDF Storage And Inference Layer API (SAIL). The source code is available online\footnote{\url{http://www.uni-koblenz-landau.de/koblenz/fb4/AGStaab/Research/systeme/NetworkedGraphs}}.  
In order to understand how does NG interacts with Sesame a closer look must be taken at Sesame's architecture, which is depicted in Figure ~\ref{fig:sesame}

Sesame's Storage And Inference Layer (SAIL) is an internal API that abstracts from the storage format used (e.g., data stored in an RDBMS, in memory, or in files -see below), and provides reasoning support over RDF triples. SAIL implementations can also be stacked on top of each other to provide other functionalities such as caching or concurrent access handling. Extensions to Sesame should be implemented as SAILs, which is the case of NG. Sesame's functional modules, such as query engines, the admin module, and RDF export, use the SAIL to perform its tasks. These functional modules can be accessed through a different API called Access API, which is composed by the Repository API and the Graph API. The Repository API provides high-level access to Sesame repositories, such as querying, storing of RDF files, extracting RDF, etc. The Graph API provides more fine-grained support for RDF manipulation (e.g.,  adding and removing individual statements). The two APIs complement each other in functionality, and are in practice often used together\footnote{\url{http://www.openrdf.org/doc/sesame/users/userguide.html/}}.
Sesame 2.3 supports three different storage formats for its repositories: in memory, in files (also called native storage), and RDBMS. Each of these formats support different maximum sizes which are not clearly defined in the documentation. For each of these storage formats there also exists the possibility of enabling either RDF entailment (by default) or RDFS entailment regime, which must be explicitly stated by the time the repository is created.

With respect to the scenarios defined in Section~\ref{sec3}, NGs are appropriate for supporting scenarios 1 (view integration), and 4 (query modularization). In Section~\ref{intro} we have already discussed on the applicability of NGs to a  view    integration scenario. 
 Example \ref{ex:example9} below shows how the query in Example \ref{ex:example8} reads in NGs syntax:
     
\begin{example} [NGs for query modularization]
 \label{ex:example9}
 \lstset{language=SPARQL,framesep=4pt,basicstyle=\scriptsize,
 showstringspaces=false, tabsize=2,numbers=none,numberstyle=\small, stepnumber=1, numbersep=5pt}
 \begin{lstlisting}
def:coleaguesView {
def:coleaguesView ng:definedBy
``CONSTRUCT {?a1 def:coleagues ?a2}
 FROM dbtune:peel
 WHERE {?a1 mo:performed ?p1 .
	?a2 mo:performed ?p2 .
	?p1 event:place ?pl1 .
	?p2 event:place ?pl1 .
	FILTER(!(?a1 = ?a2))}''^^ng:query }

# using the view in a query
SELECT ?name1 ?name2
WHERE { 
	GRAPH def:coleaguesView{
   		?artist1 def:coleagues ?artist2 } .
 	GRAPH dbtune:peel {
 		?artist1 foaf:name ?name1 .
   		?artist2 foaf:name ?name2  }}
 \end{lstlisting}\qed
\label{fig:qmodul1} 
\end{example}

\subsubsection{vSPARQL}
\label{sec4:vsparql}

Shaw et al.~\cite{Shaw2011} propose an extension to SPARQL 1.0, called vSPARQL that allows, among other features,
 to define virtual graphs and use recursive subqueries to iterate over paths of arbitrary length, including paths containing blank nodes. It also extends SPARQL by allowing to create new resources, since when developing a view, users may want to create new entities based upon the data encoded in existing datasets. vSPARQL views can be stored as intermediate results within a query but can not be stored and used in other queries. Again,   to reuse the queries they must be `copy-pasted'.

vSPARQL is implemented as patches over Jena ARQ and SDB.
Jena is a Semantic Web framework, based on Java that provides an API to extract data from, and write data  to RDF graphs. The graphs are represented as an abstract model and stored in files, databases or URIs. SDB\footnote{\url{http://openjena.org/SDB/}} is a component of Jena framework that provides storage and query of RDF datasets using relational databases.
Jena graph models can also be queried through Jena SPARQL query engine, called ARQ\footnote{\url{http://jena.sourceforge.net/ARQ/}}. 
 vSPARQL is available as a web service\footnote{\url{http://ontviews.biostr.washington.edu:8080/VSparQL_Service/}} and its source code is not available, although install instructions can be found on the web\footnote{\url{http://trac.biostr.washington.edu/trac/wiki/InstallVsparql}}.

With respect to the scenarios defined in Section ~\ref{sec3}, and due to its inability to store views definitions, vSPARQL partially supports scenarios   1 (view integration), and 4 (query modularization), as the next examples shows:  

\begin{example} [Using vSPARQL for view integration]
\label{ex:example9a}
 \lstset{language=SPARQL,framesep=4pt,basicstyle=\scriptsize,
 showstringspaces=false, tabsize=2,numbers=none,numberstyle=\small, stepnumber=1, numbersep=5pt}
 \begin{lstlisting}
SELECT DISTINCT ?artist ?record ?recordTitle
FROM dbtune:peel
FROM NAMED def:recordsView [
   CONSTRUCT {?artist foaf:made ?record}
   WHERE { 	
	{GRAPH <http://dbtune.org/jamendo/>
		{ ?artist foaf:made ?record .
		  ?artist rdf:type mo:MusicArtist .	
		  ?record rdf:type mo:Record }
	}UNION
	{GRAPH <http://dbtune.org/jamendo/>
		{ ?record foaf:maker ?artist .
		  ?artist rdf:type mo:MusicArtist .
		  ?record rdf:type mo:Record }
	}UNION
	{GRAPH <http://dbtune.org/magnatune/>
		{ ?record foaf:maker ?artist .
		  ?artist rdf:type mo:MusicArtist .
		  ?record rdf:type mo:Record }}]
WHERE { 
	GRAPH def:recordsView{ 
		?artist foaf:made ?record } .
	GRAPH dbtune:peel { 
		?record dc:title ?recordTitle }
}
 \end{lstlisting}\qed
\label{fig:qmodul1}
\end{example}

\begin{example} [Using vSPARQL for query modularization]
\label{ex:example10}

The following expression  shows how the query  in Example \ref{ex:example8}  
(scenario 4) reads in vSPARQL syntax.

 \lstset{language=SPARQL,framesep=4pt,basicstyle=\scriptsize,
 showstringspaces=false, tabsize=2,numbers=none,numberstyle=\small, stepnumber=1, numbersep=5pt}
 \begin{lstlisting}
SELECT ?name1 ?name2
FROM dbtune:peel
FROM NAMED def:coleaguesView [
	 CONSTRUCT {?a1 def:coleagues ?a2}
	 FROM dbtune:peel
	 WHERE {?a1 mo:performed ?p1 .
		?a2 mo:performed ?p2 .
		?p1 event:place ?pl1 .
		?p2 event:place ?pl1 .
		FILTER(!(?a1 = ?a2))
	}]
WHERE 
{ GRAPH def:coleaguesView{ 
   ?artist1 def:coleagues ?artist2 } .
  GRAPH dbtune:peel {
   ?artist1 foaf:name ?name1 .
   ?artist2 foaf:name ?name2  }
}

 \end{lstlisting}\qed
\label{fig:qmodul1}
\end{example}

 \subsection{Partial Support of RDF Views}

In the following  we comment on  other proposals that partially comply with our definition of RDF views, 
namely sub-queries in SPARQL, RDF indexing mechanisms, and exposing RDF views of relational databases.

\subsubsection{Support of Subqueries in SPARQL}
\label{sec4:subq}

Although  SPARQL 1.0 does not support subqueries, there exist SPARQL endpoints that have extended the language in order to allow this feature. For instance, OpenLink Virtuoso\footnote{\url{http://docs.openlinksw.com/virtuoso}} supports \texttt{SELECT} queries as part of the \texttt{WHERE} clause since version 5.  

The current working draft of SPARQL 1.1~\cite{Harris2010} includes partial support to subqueries allowing a sub-set of \texttt{SELECT} queries as part of the \texttt{WHERE} clause. These queries cannot include \texttt{FROM} or \texttt{FROM NAMED} clauses. Although SPARQL 1.1 is yet to be completed several endpoints and RDF libraries claim to support some of its incorporations, mainly subqueries. That is the case of 4store\footnote{\url{http://4store.org/}}, Jena ARQ's Fuseki\footnote{\url{http://www.openjena.org/wiki/Fuseki}}, OWLIM\footnote{\url{http://www.ontotext.com/owlim}}, and Sesame\footnote{\url{http://www.openrdf.org}} among others. 

Some authors argue on the design decisions that have been made so far, regarding subqueries, in SPARQL 1.1. In~\cite{Angles2011} the authors analyze the feasibility of using sub-queries, not only as graph patterns (within \texttt{WHERE} clause), but also as dataset clauses and as filter constraints, focusing on the definition of precise semantics and also discussing on the issues that arise related to the scope of correlated variables.

\subsubsection{RDF Indexing Mechanisms}
\label{sec4:index}
Several proposals exist aimed at enhancing SPARQL query performance  using view materialization mechanisms.  
The three approaches below  support the second scenario in Section \ref{sec3} 
(answering queries using views).

RDF-3x~\cite{Neumann2010} is an RDF triple store that implements several indexing mechanisms that lead to better query performance. It is based on a column-store persistence layer and creates in-memory indexes for each permutation of SPO objects in the datasets. They also propose a compact representation of triples.

RDFMatView~\cite{Castillo2010} proposes to build indexes over the relational representation of RDF datasets and also defines a query rewriting algorithm that allows the exploitation of this indexes by SPARQL queries. The query rewriting process is guided by a cost model that chooses between all the existent indexes,  the combination that leads to the best query execution plan. Instead of building indexes over every attribute this work proposes to carefully select which views should be materialized, but it does not provide mechanisms that assist in choosing which are the indexes that should be created. 

In~\cite{Dritsou2011} the authors define materialized views as the combination of simple path expressions over RDF graphs or shortcuts. They also propose a shortcut selection algorithm, based on linear programming, that optimizes the trade off between the expected benefit of reducing query processing cost and the space required for storing the indexes, taking into account the datasets and the query workload.

\subsubsection{Relational Data as RDF}

Several works focus on the transformation of relational data into RDF graphs\footnote{\url{http://www.w3.org/wiki/Rdb2RdfXG/StateOfTheArt}}, and in particular several tools allow exposing and querying relational data as virtual RDF graphs. 
This proliferation of tools led to a W3C working group (RDB2RDF) with the purpose of  standardizing the mapping of  relational data and relational database schemas into RDF and OWL. This group has so far
 produced several working drafts\footnote{\url{http://www.w3.org/2001/sw/rdb2rdf/}}.

D2RQ platform~\cite{Bizer2007a,Bizer2004} includes a declarative language to describe mappings between relational database schema and OWL/RDFS ontologies (D2RQ), a plug-in for the Jena and Sesame Semantic Web toolkits which translate SPARQL queries into SQL queries (D2RQ Engine) and an HTTP server that provides an SPARQL endpoint over the database (D2R Server). 

Virtuoso RDF Views~\cite{rdfviews} maps relational data into RDF and it provides a language to specify the mappings. These mappings are dynamically evaluated to create RDF graphs; consequently changes to the underlying data are reflected immediately in the RDF representation. 

Triplify~\cite{Auer2009} is another tool that focuses on publishing relational data as RDF. It uses SQL as mapping language between relational data and RDF graphs and does not provide an SPARQL endpoint. as part of the tool.

\section{Discussion and Experiments}
\label{sec5}

In Section~\ref{sec4} we have presented several RDF view specification mechanisms and study them in light of  
our definition of RDF views (Definition \ref{def:views}). From this study, it follows that  the specification mechanisms
closest to our definition are Networked Graphs, SPARQL++ and vSPARQL. We now discuss them in more detail, 
and show the results of experimental tests performed over Networked Graphs (the only implementation available at 
the time of writing this work).

\subsection{Goals}
\label{sec5:goals}

Our discussion is based on the following goals:

\begin{itemize}

\item Goal 1 ($G_{1}$): Finding out to what extent each of the three proposals supports the SPARQL 1.0 specification.

\item Goal 2 ($G_{2}$): Studying inference support under RDFs entailment.

\item Goal 3 ($G_{3}$): Assessing  scalability. The question is, how does dataset size affect  performance? Which data size restrictions apply?

\item Goal 4 ($G_{4}$): Assessing capability to integrate into or interoperate with existent Semantic Web platforms like Virtuoso, OWLIM or  Jena.
\end{itemize}

\subsubsection{Goal 1: SPARQL Support}
\label{sec5:g1}

Each of the selected RDF view specification mechanisms propose extensions to SPARQL. We want to assess to what extent they support the SPARQL  1.0 specification. Since there are differences among the SPARQL 1.0 support among different query engines and \sq\  endpoints, and some of the RDF view specification mechanisms are based on existent tools, different degrees of support could arise.

NGs are implemented over Sesame. Therefore,  the support to SPARQL is tightly coupled to the Sesame's SPARQL interpreter. 
vSPARQL also extends an existent interpreter: Jena ARQ;  thus, it should be able then to, at least, support 
the same kind of SPARQL queries supported by ARQ.  
SPARQL++,  on the contrary, implements its own SPARQL interpreter based on the translation of queries into HEX-programs. The authors prove~\cite{Polleres2007a,Polleres2007} that SPARQL++ is semantically equivalent to SPARQL, as defined in \cite{Perez2009}.

To evaluate the support of SPARQL 1.0 specification we can design a set of queries that include the most common SPARQL expressions, and use them to test the syntactic and semantic behavior of each of the mechanisms. The semantic behavior is assessed comparing the obtained results of each query, under a controlled dataset, with the expected results according to SPARQL semantics~\cite{Perez2009}. This is the approach we follow in our experiments over Networked Graphs ( Section~\ref{sec5:tests}).

\subsubsection{Goal 2: Inference Support under RDFs Entailment}
\label{sec5:g2}

SPARQL inference support under  RDFs entailment,  varies according with the different tools and implementations.
Sesame supports RDFS entailment regime as defined in the RDFS model-theoretic 
semantics~\cite{Hayes2004}. Thus, this behavior should be preserved by NGs since they are implemented as a Sesame SAIL.
ARQ also supports RDFs entailment regime, therefore vSPARQL should also support it. Finally, 
SPARQL++  does not implement RDFS entailment natively, but the inference rules presented in Section~\ref{sec2} can be represented using \texttt{CONSTRUCT} queries. As an example, Figure~\ref{fig:sparql++} shows the suggested representation for the 
 \texttt{subClass} rule presented in Section~\ref{sec2:rdf}. 

\begin{figure}[!ht]
 \lstset{language=SPARQL,framesep=4pt,basicstyle=\scriptsize,
 showstringspaces=false,
 tabsize=1,numbers=none,numberstyle=\small, stepnumber=1, numbersep=5pt}
 \begin{lstlisting}
CONSTRUCT {?A :subClassOf ?C} 
WHERE { 	?A :subClassOf ?B. ?B :subClassOf ?C. }
 \end{lstlisting}

\caption {Implementing RDFS inference support under SPARQL++ }
\label{fig:sparql++}
\end{figure}

Analogously to Goal 1, in Section \ref{sec5:tests} we show experimentally the inference support of NGs.

\subsubsection{Goal 3: Scalability}
\label{sec5:g3}

This goal has two sub-goals: (1) Assessing size limitations for each of the evaluated mechanisms; and (2) Evaluating the impact  of the dataset size  over performance. 

Although Sesame supports different types of repositories (see Section~\ref{sec4:NG}), NGs cannot be used on repositories based on RDBMS, since it only supports in-memory and (file based) native storage. This   imposes a restriction on the size of the datasets that can be used to create views. On the contrary, vSPARQL storage is implemented as patches over Jena SDB (see Section~\ref{sec4:vsparql}); thus,  it   uses relational repositories.

SPARQL++ uses DLV as its storage mechanism (see Section~\ref{sec4:sparql++}), which supports in-memory and relational storage via ODBC\footnote{\url{http://www.dlvsystem.com/dlvsystem/html/DLV_User_Manual.html}}. However, 
no precise information could be found regarding the maximum size of the datasets supported by each proposal. 
For checking sub-goal (1) we propose to locally perform load tests over different kinds of repositories.
For checking sub-goal (2) we propose to locally create repositories with different sizes, and pose a set of selected queries to measure performance. We do this for NGs  in Section \ref{sec5:tests}.

\subsubsection{Goal 4: Integration with other Platforms}
\label{sec5:g4}

This goal refers to the feasibility of integrating RDF view definition mechanisms with existent Semantic Web platforms and tools.
This integrations should be easy in the case of 
NGs and vSPARQL, since they are based on well-known platforms as Sesame and Jena. Both platforms implement a Java API that is widely used, and other tools as Virtuoso already provides connectors to interact with them\footnote{\url{http://virtuoso.openlinksw.com/dataspace/dav/wiki/Main/VOSRDFDataProviders}}
Even though, the integration of NGs to an existing Semantic Web application depends on its ability to use Sesame via its Java API. We believe that this restriction could be too strong in some contexts, mostly given that Sesame has been outperformed by other triple stores\footnote{\url{http://www.w3.org/wiki/RdfStoreBenchmarking}}.

Regarding SPARQL++, the fact that it is not based on an existent Semantic Web platform suggest that its integration with other solutions is not that straightforward. Its actual C++ implementation is intended to be used from command line, and the source code should be wrapped to give programmatic access to its functionalities.

\subsection{Experiments}
\label{sec5:tests}

\out{The  discussion above shows that,  in most of the cases, experiments   are needed to asses the fulfillment of the proposed goals. 
To perform this tests local installations of the evaluated proposals should be available, in order to control settings and datasets. As presented in Section~\ref{sec4} the source code of vSPARQL could not be obtained, and although SPARQL++ was successfully compiled and downloaded we could not overpass errors (no documentation is available). The only proposal that we could successfully compile, install and test was NG.
}

We now describe  a collection of tests aimed at  evaluating Networked Graphs with respect to  
Goals $G_1$ through $G_4$.  From the 
three proposals under study in this section, NGs is the only one whose implementation is fully available for installation, compiling, and testing. Therefore, although the design of the experiments is valid (with slight variations)  
for the three proposals, we only  report the results obtained for NGs.    
We present the dataset selection and preparation procedure,  the results obtained from the tests, and a discussion of these results.

\subsubsection{Data Selection and Preparation}
\label{sec5:datasets}

\paragraph*{Dataset Selection~~} Starting from the list of datasets published in the W3C catalogue\footnote{\url{http://www.w3.org/wiki/DataSetRDFDumps}} a selection process was performed, taking into consideration the following requirements,  closely related to our experimental goals:

\begin{itemize}

\item Requirement 1 ($R_{1}$): The  data domain should be simple enough to allow us  focusing on views problems instead of domain-related problems. This requirement  is particularly important in goals $G_1$ and $G_2$.

\item Requirement 2 ($R_{2}$): Datasets should reflect real data heterogeneity and should allow us to exemplify integration queries and problems. This requirement is highly related to goal $G_1$.

\item Requirement 3 ($R_{3}$): Datasets should be at least medium sized (over 200k triples) in order to test performance issues and scalability. This requirement applies to goal $G_3$.

\item Requirement 4 ($R_{4}$): Datasets should be available as RDF dumps, to allow using them locally. This requirement is related to all goals and refers to the ability to test local deployments of current implementations in a  controlled enivironment.

\item Requirement 5 ($R_{5}$): Datasets should include schema information in order to check  inference capabilities (at least \texttt{subClassOf} and \texttt{subPropertyOf} relationships). The fulfillment of this requirement is necessary to evaluate goal $G_2$.

\end{itemize}

In Appendix~\ref{sec:app2} we present detailed information on the datasets published by W3C and also the results of the evaluation of each requirement $R_{i}$ for each dataset $D_{j}$. Table~\ref{tab:reqSum} presents the results of the requirement evaluation, only for those datasets that fulfill most of them.   

\begin{table}[H]
\centering
\caption{Summary of the evaluation of requirements over datasets}
\label{tab:reqSum}
\begin{tabular}{|p{2cm}|p{0.6cm}|p{0.6cm}|p{0.6cm}|p{0.6cm}|p{0.8cm}|}
\hline \textbf{Dataset} & \textbf{$R_{1}$} & \textbf{$R_{2}$} & \textbf{$R_{3}$} & \textbf{$R_{4}$} & \textbf{$R_{5}$} \\ \hline
BBC John Peel & \tickYes & \tickYes & \tickYes & \tickYes & OWL \\ \hline
BTC & \tickNo & \tickYes & \tickYes & \tickYes & RDFS \\ \hline
Jamendo& \tickYes & \tickYes & \tickYes & \tickYes & OWL \\ \hline
Linked Sensor Data  & \tickYes & \tickYes & \tickYes & \tickYes & OWL \\ \hline
Magnatune & \tickYes & \tickYes & \tickYes & \tickYes & OWL \\ \hline
YAGO  & \tickYes & \tickNo & \tickYes & \tickYes & RDFS \\ \hline
\end{tabular}
\end{table}

The Billion Triple Challenge (BTC) dataset is actually a collection of datasets expressed as NQuads\footnote{\url{http://sw.deri.org/2008/07/n-quads/}} (triple plus the name of the graph) obtained by crawling Linked Data from the web. It contains data from different domains\footnote{\url{http://gromgull.net/2010/10/btc/explore.html}}, including biosciences domain data, which usually requires extra knowledge to pose meaningful queries over it.  Therefore, we consider that the BTC dataset does not completely fulfill $R_{1}$: domain understandability.
The YAGO dataset contains geographic data from different sources\footnote{\url{http://www.mpi-inf.mpg.de/yago-naga/yago/index.html}}, but it actually is the result of a consolidation and enrichment process of that data, so it does not fulfill $R_{2}$ since it does not reflect a real data integration scenario.
The datasets from the DBTune project (BBC, Jamendo and Magnatune) were the only ones that fulfilled requirements $R_{1}$ to $R_{4}$. Regarding $R_{5}$ they do not contain RDFS information inside them but refer to classes and properties defined in the MusicOntology, which is written in OWL. However,  we have extracted useful RDF schema information from the ontology using SPARQL queries based on OWL semantics \footnote{\url{http://www.w3.org/TR/owl-semantics/}}. 

Figure~\ref{fig:OWL2RDFS}  shows  the SPARQL queries used to extract schema information.

\begin{figure}[H]
 \lstset{language=SPARQL,framesep=4pt,basicstyle=\scriptsize,
 showstringspaces=false,
 tabsize=1,numbers=none,numberstyle=\small, stepnumber=1, numbersep=5pt}
 \begin{lstlisting}
CONSTRUCT {?c rdf:type rdfs:class} 
WHERE { ?c rdf:type owl:class }

CONSTRUCT {?p rdf:type rdf:Property}
WHERE { ?p rdf:type owl:DatatypeProperty}

CONSTRUCT {?p rdf:type rdf:Property}
WHERE { ?p rdf:type owl:ObjectProperty}

CONSTRUCT {?p rdf:type rdf:Property}
WHERE { ?p rdf:type owl:InverseFunctionalProperty}

CONSTRUCT {?p rdf:type rdf:Property}
WHERE { ?p rdf:type owl:TransitiveProperty}

CONSTRUCT {?p rdf:type rdf:Property}
WHERE { ?p rdf:type owl:SymmetricProperty}

CONTRUCT {?c1 rdfs:subClassOf ?c2}
WHERE {?c1 rdfs:subClassOf ?c2}

CONTRUCT {?c1 rdfs:subPropertyOf ?c2}
WHERE {?c1 rdfs:subPropertyOf ?c2}

CONSTRUCT {?p rdfs:domain ?c1}
WHERE {?p rdfs:domain ?c1}

CONSTRUCT {?p rdfs:range ?c1}
WHERE {?p rdfs:range ?c1}

 \end{lstlisting}

\caption {Extracting schema information from OWL }
\label{fig:OWL2RDFS}
\end{figure}

\paragraph*{Data Preparaton~~}

In Table~\ref{tab:datasets} we give  details about the datasets that we have used in this work.
\begin{table}[H]
\centering
\caption{Selected datasets detailed info}
\label{tab:datasets}
\begin{tabular}{|p{2cm}|p{2cm}|p{1cm}|p{1.3cm}|}
\hline \textbf{Dataset} & \textbf{Size (K Triples)} & \textbf{Size (Mb)}& \textbf{RDF syntax}\\ \hline
BBC J.Peel & $\sim$ 380 & 22 & XML\\ \hline
Jamendo & $\sim$ 1000 & 57 & XML\\ \hline
Magnatune & $\sim$ 600 & 36 & XML\\ \hline
MusicOntology & $\sim$ 1 & 0.07 & N3\\ \hline
\end{tabular}
\end{table}

To evaluate the effects of the number of triples over performance, original datasets were split into smaller files\footnote{Aprox. 100.000 lines of text in each file} and three different datasets $DT_{i}$ were created. Table~\ref{tab:datasplit} reports the size of each dataset.

\begin{table}[H]
\centering
\caption{Sub-datasets}
\label{tab:datasplit}
\begin{tabular}{|p{1.2cm}|p{2.5cm}|p{2cm}|}
\hline
\textbf{Dataset} & \textbf{Size (K Triples)} &\textbf{Size (Mb)} \\ \hline
$DT_{1}$ &  $\sim$ 500 &  28.8 \\ \hline
$DT_{2}$ &  $\sim$ 1000 & 57.5 \\ \hline
$DT_{3}$ &  $\sim$ 2000 & 115 \\ \hline
\end{tabular}
\end{table}

The   datasets were loaded in different Sesame repositories. We have created 8 repositories with the following characteristics:
\begin{itemize}
\item In-memory storage without RDFS entailment support ($MEM_i$ i=1 to 3)
\item Sesame native storage without RDFS entailment support ($NAT_i $ i=1 to 3)
\item Sesame native storage with RDFS entailment support ($NATR$ and $NATR_1$)
\end{itemize}

Each of the repositories described above has been loaded with its correspondent set of triples $DT_i$, except $NATR$ which used in Test 2. For instance, we have the $MEM_1, NAT_1$ and $NATR_1$ repositories populated  with dataset $DT_1$.  The contents of repository $NATR$ will be described later, in Section \ref{sec5:exp2}

\subsubsection{Experimentation Details}
\label{sec5:pertests}

Our tests were run on a desktop PC ( 2.53 GHz Intel Core 2 Duo, 2 Gb RAM) under the Ubuntu 10 operating system. Sesame 2.3 server was installed under Apache Tomcat server (version 6.0.32).Default Tomcat settings have been changed to increase heap size to 1Gb.
We now describe the tests performed, aimed at evaluating NG's compliance with goals $G_1$ through $G_4$. 
For each one of them we provide the queries and details on the datasets and repositories, and report the results of the experiments.

\paragraph {Test 1: SPARQL Support}
\label{sec5:exp1}

The purpose of this test was to check to what extent Networked Graphs support SPARQL 1.0 specification.
The test consisted of the following steps:

\begin{enumerate}
\item Design a set of \texttt{CONSTRUCT} queries $Q_i$ covering most of SPARQL functionalities;
\item For each of the $Q_i$ queries defined in step 1:
\begin{enumerate}[label*=\arabic*.]
\item Build the NG $NG_i$ defined by query $Q_i$;
\item Run $Q_i$;
\item Run \texttt{SELECT * FROM $NG_i$ WHERE \{?s ?p ?o\}};
\item Compare the results of both runs and enumerate  the differences, if any. Identical results of 2.2 and 2.3 imply SPARQL compliance.
\end{enumerate}
\end{enumerate}

\paragraph*{Datasets~~} The focus of this test was on the semantics of the \texttt{CONSTRUCT} queries in Sesame and NGs behaviour. Thus, we only used repository $MEM_1$ (we do not care here about RDFS entailment and performance).

\paragraph*{Queries~~} We now  describe the set of queries performed in this test. The queries combine different SPARQL clauses (presented in Section~\ref{sec2:sparql}) adding functionalities incrementally. They are organized in the following groups:

\begin{itemize}
\item \textbf{Group A}: Queries that only have a graph pattern. One query for each  possible graph pattern (BGP, group pattern, optional pattern, union pattern and patterns on named graphs),
\item \textbf{Group B}: Queries obtained by adding \texttt{FILTER} expressions to queries in Group A,
\item \textbf{Group C}: Queries obtained by adding negation clauses to queries in Group B,
\item \textbf{Group D}: Queries obtained by adding \texttt{ORDER BY} clauses to queries in Group C,
\end{itemize}
 
Appendix ~\ref{sec:appExperiments} gives a  detail of   the  queries used in the experiments. Table~\ref{tab:queries} summarizes the queries in each group  for  further referencing them in the remainder  of this section.

\begin{table}[H]
\caption{Queries in each group for test 1}
\centering
\begin{tabular}{|p{2cm}|p{0.5cm}|p{0.5cm}|p{0.5cm}|p{0.5cm}|}
\hline
& A & B & C & D\\ \hline
BGP &$q_1$ & $q_6$ & $q_{11}$ & $q_{15}$ \\ \hline
Group GP & $q_2$ & $q_7$ & $q_{12}$ & $q_{16}$ \\ \hline
Optional GP & $q_3$ & $q_8$ & $q_{13}$ & $q_{17}$ \\ \hline
Union GP & $q_4$ & $q_9$ & $q_{14}$ & $q_{18}$ \\ \hline
Graph FROM NAMED & $q_5$ & $q_{10}$ & \tickNo  & \tickNo \\ \hline
\end{tabular}
\label{tab:queries}
\end{table}

\paragraph {Results}
\label{sec5:resExp1}

The results  obtained show that only query $q_{10}$ (which contains a \texttt{FILTER} expression combined with \texttt{GRAPH} expressions) does not retrieve the expected results, neither as a \texttt{CONSTRUCT} query in Sesame, 
nor as a view definition using NGs. Due to this observation these kinds of queries were not included 
in groups C and D.

\out{\begin{table}[htbp]
\caption{Results of test 1}
\centering
\begin{tabular}{|l|l|l|}
\hline
\multicolumn{ 1}{|c|}{group A} & q1 & \tickYes \\ \cline{ 2- 3}
\multicolumn{ 1}{|l|}{} & q2 & \tickYes \\ \cline{ 2- 3}
\multicolumn{ 1}{|l|}{} & q3 & \tickYes \\ \cline{ 2- 3}
\multicolumn{ 1}{|l|}{} & q4 & \tickYes \\ \cline{ 2- 3}
\multicolumn{ 1}{|l|}{} & q5 & \tickYes \\ \hline
\multicolumn{ 1}{|c|}{group B} & q6 & \tickYes \\ \cline{ 2- 3}
\multicolumn{ 1}{|l|}{} & q7 & \tickYes \\ \cline{ 2- 3}
\multicolumn{ 1}{|l|}{} & q8 & \tickYes \\ \cline{ 2- 3}
\multicolumn{ 1}{|l|}{} & q9 & \tickYes \\ \cline{ 2- 3}
\multicolumn{ 1}{|l|}{} & q10 & \tickNo \\ \hline
\multicolumn{ 1}{|c|}{group C} & q11 & \tickYes \\ \cline{ 2- 3}
\multicolumn{ 1}{|l|}{} & q12 & \tickYes \\ \cline{ 2- 3}
\multicolumn{ 1}{|l|}{} & q13 & \tickYes \\ \cline{ 2- 3}
\multicolumn{ 1}{|l|}{} & q14 & \tickYes \\ \hline
\multicolumn{ 1}{|c|}{group D} & q15 & \tickYes \\ \cline{ 2- 3}
\multicolumn{ 1}{|c|}{} & q16 & \tickYes \\ \cline{ 2- 3}
\multicolumn{ 1}{|l|}{} & q17 & \tickYes \\ \cline{ 2- 3}
\multicolumn{ 1}{|l|}{} & q18 & \tickYes \\ \hline
\end{tabular}
\label{tab:resExp1}
\end{table}
}

\paragraph {Test 2: RDFS Inference Support} 
\label{sec5:exp2}

The purpose of this test is  to check to what extent Networked Graphs support RDFS entailment regime.
The test consisted of the following steps:
\begin{enumerate}
\item Build a simple dataset that allows us to control the results of the application of RDFS rules presented in Section~\ref{sec2:rdf};
\item Load the dataset in repository $NATR$;
\item Design a set of \texttt{CONSTRUCT} queries $I_i$ for testing each of the rules;
\item For each of the queries $I_i$ defined in step 2:
\begin{enumerate}[label*=\arabic*.]
\item Build an NG $NG_i$ defined by query $I_i$ in repository $NATR$;
\item Run \texttt{SELECT * FROM $NG_i$} \texttt{WHERE \{?s ?p ?o\}};
\item Compare obtained results with expected results under RDFS entailment (see Table~\ref{tab:RDFSexp})
\end{enumerate}
\end{enumerate}

\paragraph*{Datasets~~} We built a very simple dataset that provided us with a controlled environment for checking
 RDFS entailment rules. The triples contained in this simple dataset are the following (prefix clauses are omitted for the sake of readability):

\begin{mydataset}
 \lstset{language=SPARQL,framesep=4pt,basicstyle=\scriptsize, showstringspaces=false, tabsize=1,numbers=none}
 \begin{lstlisting}
dat:inferenceTest {
 mo:singer rdfs:subPropertyOf mo:performer .
 mo:performer rdfs:subPropertyOf eve:agent .
 dat:JohnnyCash mo:singer dat:PersonalJesus .
 mo:Record rdfs:subClassOf 
 	                mo:MusicalManifestation .
 mo:LiveAlbum rdfs:subClassOf mo:Record
 dat:TheManComesAround rdf:type mo:Record .
 mo:chart_position rdfs:domain 
 	                mo:MusicalManifestation .
 dat:IWalkTheLine mo:chart_position ``1'' .
 mo:recorded rdfs:range mo:Record .
 dat:JohnnyCash mo:recorded 
 	dat:AmericanRecordings .		
} 
 \end{lstlisting}
\label{dat:rdfsDat}
\end{mydataset}

\paragraph*{Queries~~}  We have designed one query for each of the RDFS rules presented in Section~\ref{sec2:rdf}. The following query set contains each of the designed queries. Tables~\ref{tab:RDFexp} and \ref{tab:RDFSexp} presents their expected results under RDF and RDFS entailment, respectively.

\begin{myqueryset}
\label{qe:qInf}
 \lstset{language=SPARQL,framesep=4pt,basicstyle=\scriptsize, showstringspaces=false, tabsize=1,numbers=none}
 \begin{lstlisting}
 
#subProperty (1)
CONSTRUCT {?p rdfs:subPropertyOf event:agent}
WHERE {?p rdfs:subPropertyOf event:agent}

#subProperty (2)
CONSTRUCT {dat:JohnnyCash mo:performer ?p}
WHERE {dat:JohnnyCash mo:performer ?p}

#subClass (3)
CONSTRUCT {?p rdfs:subClassOf 
           mo:MusicalManifestation}
WHERE {?p rdfs:subClassOf 
           mo:MusicalManifestation}

#subClass (4)
CONSTRUCT {dat:TheManComesAround rdf:type ?p}
WHERE {dat:TheManComesAround rdf:type ?p}

#typing (5)
CONSTRUCT {dat:IWalkTheLine rdf:type ?p}
WHERE {dat:IWalkTheLine rdf:type ?p}

#typing (6)
CONSTRUCT {dat:AmericanRecordings rdf:type ?p}
WHERE {dat:AmericanRecordings rdf:type ?p}

 \end{lstlisting}

\end{myqueryset}

\begin{table}
\caption{Expected results for queries in Test 2 without RDFS entailment regime}
\centering
\begin{tabular}{|l|}
\hline
\textbf{subPropertyOf (1)} \\ \hline
mo:performer rdfs:subPropertyOf event:agent \\ \hline
\textbf{subPropertyOf (2)} \\ \hline
empty\\ \hline
\textbf{subClassOf (3)} \\ \hline
mo:Record rdfs:subClassOf mo:MusicalManifestation \\ \hline
\textbf{subClassOf (4)} \\ \hline
dat:TheManComesAround a mo:Record \\ \hline
\textbf{typing (5)} \\ \hline
empty\\ \hline
\textbf{typing (6)} \\ \hline
empty\\ \hline
\end{tabular}
\label{tab:RDFexp}
\end{table}

\begin{table}
\caption{Expected results for queries in Test 2 under RDFS entailment regime}
\centering
\begin{tabular}{|l|}
\hline
\textbf{subPropertyOf (1) }\\ \hline
mo:performer rdfs:subPropertyOf event:agent \\ \hline
mo:singer rdfs:subPropertyOf event:agent \\ \hline
\textbf{subPropertyOf (2)} \\ \hline
dat:JohnnyCash mo:performer dat:PersonalJesus \\ \hline
\textbf{subClassOf (3) }\\ \hline
mo:Record rdfs:subClassOf mo:MusicalManifestation \\ \hline
mo:LiveAlbum rdfs:subClassOf mo:MusicalManifestation \\ \hline
\textbf{subClassOf (4)} \\ \hline
dat:TheManComesAround rdf:type mo:Record \\ \hline
dat:TheManComesAround rdf:type mo:MusicalManifestation \\ \hline
\textbf{typing (5)} \\ \hline
dat:IWalkTheLine rdf:type mo:Record \\ \hline
dat:IWalkTheLine rdf:type mo:MusicalManifestation \\ \hline
\textbf{typing (6) }\\ \hline
dat:AmericanRecordings rdf:type mo:Record \\ \hline
dat:AmericanRecordings rdf:type mo:MusicalManifestation \\ \hline
\end{tabular}
\label{tab:RDFSexp}
\end{table}

\paragraph*{Results~~} 
\label{sec5:resExp2}

For every query in Test 2 the obtained results correspond to those expected under RDFS entailment regime, presented in Table \ref{tab:RDFSexp} .

\paragraph {Test 3: Scalability}
\label{sec5:exp3}

The purpose of this test is two-fold: (1) To asses size-limitations for each of the repositories supported by NG; and (2) To evaluate the impact that datasets size has over performance. To asses size-limitations for in-memory and native repositories we have loaded triples incrementally until errors where obtained.
To evaluate the impact of datasets size over performance we have gone through the following steps:

\begin{enumerate}
\item For each repository $MEM_i$, $NAT_i$ and $NATR_1$ described above:
\begin{enumerate}[label*=\arabic*.]
\item For each of the queries in Test 1 $Q_i$, 
\begin{enumerate}[label*=\arabic*.]
\item Build the NG $NG_i$ defined by query $Q_i$;
\item Run \texttt{SELECT * FROM $NG_i$} \texttt{WHERE \{?s ?p ?o\}};
\item Measure the  execution time.
\end{enumerate}
\end{enumerate}
\end{enumerate}

\paragraph*{Datasets~~} We used the datasets described in  Table ~\ref{tab:datasplit}.
\paragraph*{Queries~~}  The queries used in this test are the same queries  presented in Test 1.

\paragraph {Results}
\label{sec5:resExp3}

Regarding size limitations our tests show that, with our configuration, in-memory repositories support loading at most 400Mb in RDF-XML format,  which represents 7 million triples approximately.
Under the same conditions we were able to load up to 1Gb into a Sesame native repository, which represents 20 million triples approximately.

Before presenting results on performance tests we want to point out that queries that use UNION graph pattern show a very poor performance under our configuration (i.e: query $q_4$ was still running after 1 hour on repository $MEM_2$). Due to this, we have excluded these kinds of queries from our tests.

\begin{figure*}[t!]
  \centering
  \subfloat[Native repositories results, aggregated by queries groups.]{\label{fig:natGroup}\includegraphics[width=0.5\textwidth]{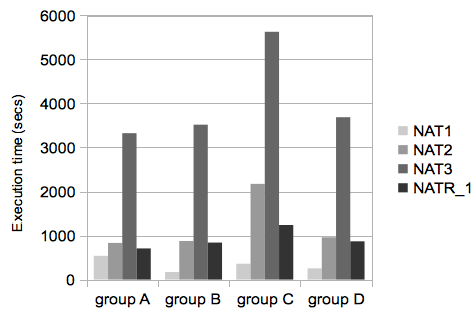}}                
  \subfloat[In memory repositories results, aggregated by queries groups.]{\label{fig:memGroup}\includegraphics[width=0.5\textwidth]{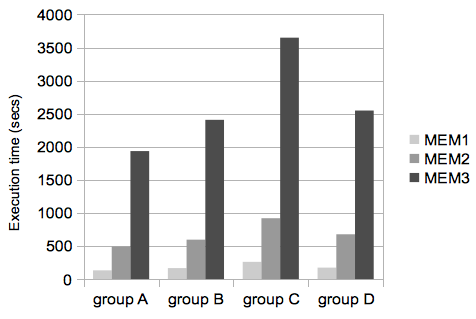}}
  
  \subfloat[Native repositories results, aggregated by query type.]{\label{fig:natType}\includegraphics[width=0.5\textwidth]{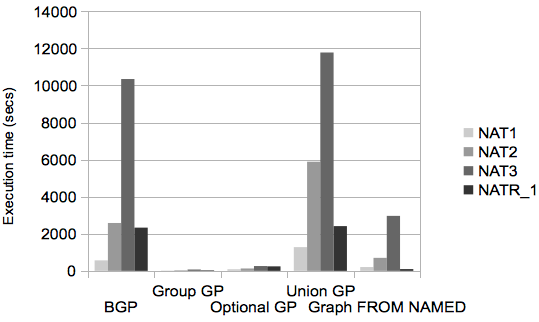}}
  \subfloat[In memory repositories results, aggregated by query type.]{\label{fig:memType}\includegraphics[width=0.5\textwidth]{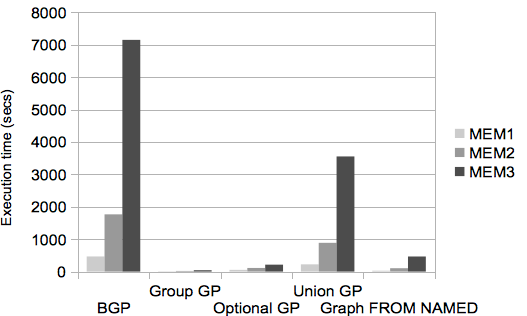}}
  \caption{Results from Test 3}
  \label{fig:gresExp3}
\end{figure*}

Table~\ref{tab:exp3nat0} presents, for each of the $NG_i$ defined, its execution time over Sesame native repositories $NAT_1$, $NAT_2$, $NAT_3$ and $NATR_1$. The last one has RDFS inference capabilities. 
Table~\ref{tab:exp3mem0} presents, for each of the $NG_i$ defined, its execution time over Sesame in-memory repositories $MEM_1$, $MEM_2$ and $MEM_3$.

\begin{table}[H]
\centering
\caption{Execution time (in seconds) for each query over Sesame native repositories}
\begin{tabular}{|l|r|r|r|r|}
\hline
\textbf{Query} & \textbf{$NAT_1$} & \textbf{$NAT_2$} & \textbf{$NAT_3$} & \textbf{$NATR_1$} \\ \hline
$NG_1$ & 590 & 2489 &  10056 & 2475\\ \hline
$NG_2$ & 3 & 10 &  24 & 11 \\ \hline
$NG_3$ & 63 & 128 &  256 & 248\\ \hline
$NG_4$ & 1839 & N/A & N/A  & N/A\\ \hline
$NG_5$ & 204 & 702 &  2965 & 94\\ \hline
$NG_6$ & 355 & 2474 &  10225 & 2242\\ \hline
$NG_7$ & 19 & 33 &  76 & 37\\ \hline
$NG_8$ & 142 & 124 &  256 & 241 \\ \hline
$NG_{11}$ & 637 & 2645 &  10406 & 2269 \\ \hline
$NG_{12}$ & 17 & 36 &  79 & 37\\ \hline
$NG_{13}$ & 64 & 128 &  265 & 241\\ \hline
$NG_{14}$ & 720 & 5886 &  11777 & 2413\\ \hline
$NG_{15}$ & 678 & 2714 &  10696 & 2334\\ \hline
$NG_{16}$ & 19 & 37 &  103 & 34\\ \hline
$NG_{17}$ & 69 & 131 &  268 & 240\\ \hline
\end{tabular}
\label{tab:exp3nat0}
\end{table}

\begin{table}
\centering
\caption{Execution time (in seconds) for each query over Sesame in-memory repositories}
\begin{tabular}{|r|r|r|r|}
\hline
\textbf{Query}& \textbf{$MEM_1$} & \textbf{$MEM_2$} & \textbf{$MEM_3$} \\ \hline
$NG_1$ & 446 & 1757 & 7057 \\ \hline
$NG_2$ & 3 & 6 & 13 \\ \hline
$NG_3$ & 52 & 105 & 203 \\ \hline
$NG_4$ & 0 & 0 & 0 \\ \hline
$NG_5$ & 27 & 104 & 465 \\ \hline
$NG_6$ & 434 & 1666 & 6974 \\ \hline
$NG_7$ & 11 & 22 & 46 \\ \hline
$NG_8$ & 53 & 100 & 202 \\ \hline
$NG_{11}$ & 533 & 1773 & 7208 \\ \hline
$NG_{12}$ & 12 & 23 & 61 \\ \hline
$NG_{13}$ & 53 & 109 & 220 \\ \hline
$NG_{14}$ & 446 & 1774 & 7106 \\ \hline
$NG_{15}$ & 454 & 1869 & 7367 \\ \hline
$NG_{16}$ & 12 & 28 & 49 \\ \hline
$NG_{17}$ & 55 & 135 & 227 \\ \hline\end{tabular}
\label{tab:exp3mem0}
\end{table}

Tables ~\ref{tab:exp3nat1} and~\ref{tab:exp3mem1} present the results in Tables~\ref{tab:exp3nat0} and~\ref{tab:exp3mem0}, respectively, aggregated by the query groups presented in Section~\ref{sec5:exp1}.

\begin{table}
\centering
\caption{Average execution time (in seconds) for each group of queries over Sesame native repositories}
\begin{tabular}{|l|r|r|r|r|}
\hline
 & \textbf{$NAT_1$} & \textbf{$NAT_2$} &\textbf{$NAT_3$}& \textbf{$NATR_1$} \\ \hline
\textbf{group A} & 539.80 & 832.25 & 3325.25 &  707.00 \\ \hline
\textbf{group B} & 172.00 & 877.00 & 3519.00 &   840.00\\ \hline
\textbf{group C} & 359.50 & 2173.75 & 5629.25 &   1240.00\\ \hline
\textbf{group D} & 255.33 & 960.67 & 3689.00 &   869.33\\ \hline
\end{tabular}
\label{tab:exp3nat1}
\end{table}

\begin{table}
\centering
\caption{Average execution time (in seconds) for each group of queries over Sesame in-memory repositories}
\begin{tabular}{|l|r|r|r|}
\hline
 & \textbf{$MEM_1$} & \textbf{$MEM_2$} & \textbf{$MEM_3$} \\ \hline
\textbf{group A} & 132.00 & 493.00 & 1934.50 \\ \hline
\textbf{group B} & 166.00 & 596.00 & 2407.33 \\ \hline
\textbf{group C} & 261.00 & 919.75 & 3648.75 \\ \hline
\textbf{group D} & 173.67 & 677.33 & 2547.67 \\ \hline
\end{tabular}
\label{tab:exp3mem1}
\end{table}

Tables ~\ref{tab:exp3nat2} and~\ref{tab:exp3mem2} present the results shown in Tables~\ref{tab:exp3nat0} and~\ref{tab:exp3mem0}, respectively,  aggregated according to the kind of graph pattern used in each query (see Table \ref{tab:queries}). Figure \ref{fig:gresExp3}   presents the graphs corresponding to the results in Tables\ref{tab:exp3nat1},\ref{tab:exp3mem1},\ref{tab:exp3nat2} and \ref{tab:exp3mem2}. 

\begin{table}[H]
\centering
\caption{Average execution time (in seconds) for queries, organized by feature, over Sesame native repositories}
\begin{tabular}{|p{2cm}|p{1cm}|p{1cm}|p{1cm}|p{1cm}|}
\hline
 & \textbf{$NAT_1$} & \textbf{$NAT_2$} &\textbf{$NAT_3$}& \textbf{$NATR_1$} \\ \hline
\textbf{BGP} & 565.00 & 2580.50 & 10345.75 &  2330.00\\ \hline
\textbf{Group GP} & 14.50 & 29.00 & 70.50 &  29.75\\ \hline
\textbf{Optional GP} & 84.50 & 127.75 & 258.75 &  242.50\\ \hline
\textbf{Union GP} & 1279.50 & 5886.00 & 11777.00 &   2413.00\\ \hline
\textbf{Graph FROM NAMED} & 204.00 & 702.00 & 2965.00 &  94.00\\ \hline
\end{tabular}
\label{tab:exp3nat2}
\end{table}

\begin{table}
\centering
\caption{Average execution time (in seconds) for queries, organized by feature, over Sesame in-memory repositories}
\begin{tabular}{|p{2cm}|p{1cm}|p{1cm}|p{1cm}|}
\hline
 & \textbf{$MEM_1$} & \textbf{$MEM_2$} & \textbf{$MEM_3$} \\ \hline
\textbf{BGP} & 466.75 & 1766.25 & 7151.50 \\ \hline
\textbf{Group GP} & 9.50 & 19.75 & 42.25 \\ \hline
\textbf{Optional GP} & 53.25 & 112.25 & 213.00 \\ \hline
\textbf{Union GP} & 223.00 & 887.00 & 3553.00 \\ \hline
\textbf{Graph FROM NAMED} & 27.00 & 104.00 & 465.00 \\ \hline
\end{tabular}
\label{tab:exp3mem2}
\end{table}

\subsection{Results discussion}
\label{sec5:results}

The results obtained in test 1 allow us to state, regarding goal $G_1$, that NGs support of SPARQL 1.0 specification is actually restrained to Sesame's support of this query language. It allows to build quite complex queries, although we have noticed that \texttt{CONSTRUCT} queries that combine \texttt{FILTER} and \texttt{GRAPH} expressions do not behave as expected. For example, the query presented in Example~\ref{fig:constructNo} returns an empty graph, although the query in Example~\ref{fig:constructOk} returns several triples and there are artists whose name contains the string ``the''.

\begin{example} 
\label{ex:exampleNo}
 \lstset{language=SPARQL,framesep=4pt,basicstyle=\scriptsize,
 showstringspaces=false, tabsize=1,numbers=none,numberstyle=\small, stepnumber=1, numbersep=5pt}
 \begin{lstlisting}
 
CONSTRUCT {?name foaf:made ?work}
FROM NAMED <http://dbtune.org/magnatune>
WHERE
{ GRAPH <http://dbtune.org/magnatune> {
    ?work foaf:maker ?artist .
    ?artist foaf:name ?name .
    FILTER (REGEX(str(?name), ``^The'',``i''))
  }
}

 \end{lstlisting} \qed
\label{fig:constructNo}
\end{example}

\begin{example} 
\label{ex:exampleOk}
 \lstset{language=SPARQL,framesep=4pt,basicstyle=\scriptsize,
 showstringspaces=false, tabsize=1,numbers=none,numberstyle=\small, stepnumber=1, numbersep=5pt}
\begin{lstlisting}

CONSTRUCT {?name foaf:made ?work}
FROM NAMED <http://dbtune.org/magnatune>
WHERE
{ GRAPH <http://dbtune.org/magnatune> {
    ?work foaf:maker ?artist .
    ?artist foaf:name ?name }
}
\end{lstlisting}\qed
\label{fig:constructOk}
\end{example}

Regarding goal $G_2$, our tests show that NGs behaviour is consistent with RDFS entailment regime, supporting all the rules presented in Section~\ref{sec2:rdf}.

Regarding goal $G_3$, according to our tests, NGs have strong restrictions regarding the maximum size of repositories. Performance tests show that some queries (those that contain \texttt{UNION} expressions like $NG_4$), although supported by NG, are impractical since we obtained response times of several hours for rather small datasets. The comparison of overall performance of in-memory vs native repositories shows that, as expected, in-memory repositories have better response times (see Figure~\ref{fig:gresExp3}).
Results also show that performance degrades with the size of the datasets, in a way such that the degradation rate observed in native repositories is higher than in memory repositories.
The experiments performed over a native repository with RDFS inference capabilities shows that enabling this feature also degrades performance. The comparison of  the results obtained over different repositories shows that the degradation in performance leads repository $NATR_1$ to behave similarly to repository $NAT_2$, which has twice the amount of data loaded. Furthermore, in Table~\ref{tab:exp3nat0} we can see that response time in $NATR_1$ is, on average, 4 times grater than response time in $NAT_1$.

\section{Conclusions and Open Research Directions}
\label{sec6}

In this work we  have reviewed existent work on views over RDF datasets, and   discussed the application of existent view definition mechanisms to four scenarios in which views have proved to be useful in traditional (relational) data management systems. To give a  framework for the  discussion we provided a definition of views over RDF datasets, an issue over which there is  no consensus so far.
We finally  chose the three proposals  closer to this definition, and analyzed them with respect to four selected goals.
 
Let us recall the four scenarios presented in Section~\ref{sec3}: virtual data integration, query answering using views, data security, and query modularization. From our study, it follows that  for each of these scenarios, the ability to support views over RDF datasets as stated  in Definition ~\ref{def:views}  could be relevant in the context of Semantic Web. Let us further comment on this.  Regarding \emph{virtual data integration}, the ability to dynamically define, store and reuse RDF graphs  provided by Networked Graphs~\cite{Schenk2008}, allows 
us to  query heterogeneous data sources, as the examples in Section~\ref{intro}  (illustrating the application of NGs to this scenario) show. 
We also showed  that in the Semantic Web context, existent work on the  \emph{query answering using views} scenario,  is mostly related to indexing and query optimization. Some approaches focus on optimizing access to ``Subject,Predicate,Object'' permutations, like RDF-3x~\cite{Neumann2010}, whereas other works are aimed at materializing specific queries (e.g., RDFMatView~\cite{Castillo2010}) or path expressions (e.g.,~\cite{Dritsou2011}). These materialized queries and path expressions are then used by the query evaluation 
system to optimize user queries. However, no mechanisms are provided to allow the user to define and store those views.
We also commented in Section \ref{sec3} that named graphs   have   been proved useful to specify data access policies and \emph{data security} by means of specifying control access permisions~\cite{Flouris2010}.  This suggests that the capability to define views proposed in the present work could be relevant in this scenario (since a named graphs is  actually a kind of view).
Finally, regarding \emph{query modularization},  in Sections~\ref{sec3} and~\ref{sec4} we have also presented examples on the usefulness of views in this context, by showing how the former can be implemented to enhance query modularization in the proposals we have studied. Again, these proposals however, do not fully implement our approach to what a view over RDF data should be.

We performed tests over Networked Graphs since, by the time of writing this work, it was the only tool that could be fully downloaded, compiled, installed and used. However,  the  tests can be performed to evaluate other proposals.  The experimental results, presented in Section~\ref{sec5:resExp1}, show that  is feasible to use NGs, although, some issues arise. The more relevant  of them are: (1) Restrictions apply to the kinds of queries that can be answered within a real user-compatible time (UNION queries have very bad performance compared with other queries); (2) Query performance degrades on average more than 10 times when comparing datasets of 500 K triples vs datasets of 2000 K triples; and (3) Query performance degrades on average 4 times when comparing datasets of 500 K triples with and without RDFS inference support.

\subsection{Open Issues}

A question that arises from our study refers to whether or not a mechanism to explicitly define RDF views in the SPARQL specification is needed. Even though there is no sign that  this issue is currently under consideration, we believe that including such mechanism  like, for instance, a \texttt{CREATE VIEW} statement, 
would allow to simplify queries, and also facilitate producing  a well-defined semantics to tackle other issues (for instance query rewriting).
Although under a different data model, this and other several issues on views have been already discussed during the early stages of XML~\cite{Abiteboul1999}.

Other open issues are those related to the optimization of query  execution plans when the query includes  one or more views. As stated in~\cite{Schmidt2010} JOIN operations implemented as AND are among the main source of complexity in SPARQL fragments without OPTIONAL clauses. Actual implementations of views, like NGs, do not provide mechanisms to optimize the execution plan for queries including
views. If a query uses an NG,  the query that defines this NG is first posed to retrieve triples and then, these triples are used in the outer query. Mechanisms for explicitly defining views may allow query rewriting techniques to be applied, as it has been traditionally done in database systems. These rewriting techniques should aim at minimizing query execution costs, both in terms of size and time, for instance: optimizing join operations and filtering triples as soon as possible.

Finally, and regarding materialized views, none of the existing approaches deals with RDF materialized views update and maintenance. These issues, particularly important in the Semantic Web setting due to the dynamic nature of web data, requires the   attention of the research community.


\bibliographystyle{spmpsci}      


\appendix

\section{Queries}
\label{sec:appExperiments}

In this appendix we give details on the queries presented in Section~\ref{sec5:exp1}. For each one of them we present the SPARQL \texttt{CONSTRUCT} query used to define the NG and also provide a description of the query results. Prefix clauses are omitted in order to facilitate the reading.

\subsection*{Group A: queries only with \texttt{WHERE} clauses} 

\begin{myquery}
Artists and the records they have made
 \lstset{language=SPARQL,framesep=4pt,basicstyle=\scriptsize, showstringspaces=false,
 tabsize=1,numbers=none,numberstyle=\small, stepnumber=1, numbersep=5pt}
 \begin{lstlisting}
 # simple BGP

CONSTRUCT {?artist foaf:made ?record}

WHERE{
	?artist a mo:MusicArtist .
	?record a mo:Record .
	?record foaf:maker ?artist .
	?artist foaf:name ?name
}
 \end{lstlisting}

\label{qe:q1}
\end{myquery}

\begin{myquery}
Artists and their performances, where the performance has been recorded and published as a track with a track number.
 \lstset{language=SPARQL,framesep=4pt,basicstyle=\scriptsize, showstringspaces=false,
 tabsize=1,numbers=none,numberstyle=\small, stepnumber=1, numbersep=5pt}
\begin{lstlisting}
# group graph pattern

CONSTRUCT {?artist mo:performed ?performance }

WHERE{
	{?performance mo:performer ?artist}
	{?performance mo:recorded_as ?signal}
	{?signal mo:published_as ?track}
	{?track mo:track_number ?num}
}

\end{lstlisting}
\label{qe:q2}
\end{myquery}

\begin{myquery}
Artists and their name. If available, also retrieves images of the artist, biographic information, other entries that represent the same artist and location of the artist
 \lstset{language=SPARQL,framesep=4pt,basicstyle=\scriptsize, showstringspaces=false,
 tabsize=1,numbers=none,numberstyle=\small, stepnumber=1, numbersep=5pt}
\begin{lstlisting}
# optional graph pattern

CONSTRUCT {?artist foaf:name ?name; 
           foaf:img ?img;
           mo:biography ?bio; 
           bio:olb ?olb;
           owl:sameAs ?artist2;
           foaf:based_near ?p }
           
WHERE {
			?artist a mo:MusicArtist ;
			foaf:name ?name .
			
OPTIONAL { ?artist foaf:img ?img}.
OPTIONAL { ?artist mo:biography ?bio}.
OPTIONAL { ?artist bio:olb ?olb}.
OPTIONAL { ?artist owl:sameAs ?artist2 }.
OPTIONAL { ?artist foaf:based_near ?p }
}

\end{lstlisting}
\label{qe:q3}
\end{myquery}

\begin{myquery}
Artist and records, where the artist has made the record or the record was made by the artist.
 \lstset{language=SPARQL,framesep=4pt,basicstyle=\scriptsize, showstringspaces=false,
 tabsize=1,numbers=none,numberstyle=\small, stepnumber=1, numbersep=5pt}
\begin{lstlisting}
# union graph pattern

CONSTRUCT {?artist foaf:made ?record}
WHERE{
	{?artist a mo:MusicArtist .
	?record a mo:Record .
	?record foaf:maker ?artist } 
	UNION
	{?artist a mo:MusicArtist .
	?record a mo:Record .
	?artist foaf:made ?record }
}

\end{lstlisting}
\label{qe:q4}
\end{myquery}

\begin{myquery}
Artist and works, where the artist has made the work in Jamendo dataset or the work has been made by the artist in Magnatune dataset
 \lstset{language=SPARQL,framesep=4pt,basicstyle=\scriptsize, showstringspaces=false,
 tabsize=1,numbers=none,numberstyle=\small, stepnumber=1, numbersep=5pt}
\begin{lstlisting}
#graph pattern applied to a named graph
CONSTRUCT {?artist1 foaf:made ?work1 .
           ?artist2 foaf:made ?work2}
FROM NAMED <http://dbtune.org/jamendo>
FROM NAMED <http://dbtune.org/magnatune>
WHERE
{  GRAPH <http://dbtune.org/jamendo>{
    ?artist1 foaf:made ?work1 } .
  GRAPH <http://dbtune.org/magnatune> {
	?work2 foaf:maker ?artist2 }}

\end{lstlisting}
\label{qe:q5}
\end{myquery}

\subsection*{Group B: queries in Group A plus \texttt{FILTER} expressions} 

\begin{myquery}
Artists and the records they have made, only for artists which name begins with ``the''
 \lstset{language=SPARQL,framesep=4pt,basicstyle=\scriptsize, showstringspaces=false,
 tabsize=1,numbers=none,numberstyle=\small, stepnumber=1, numbersep=5pt}
\begin{lstlisting}
# q1 plus FILTER condition
CONSTRUCT {?artist foaf:made ?record}
WHERE{
	?artist a mo:MusicArtist .
	?record a mo:Record .
	?record foaf:maker ?artist .
	?artist foaf:name ?name .
	FILTER (REGEX(str(?name), ``^the'', ``i''))}
\end{lstlisting}
\label{qe:q6}
\end{myquery}

\begin{myquery}
Artists and their performances, where the performance has been recorded and published as a track with a track number, and the track number is between 1 and 5
 \lstset{language=SPARQL,framesep=4pt,basicstyle=\scriptsize, showstringspaces=false,
 tabsize=1,numbers=none,numberstyle=\small, stepnumber=1, numbersep=5pt}
\begin{lstlisting}
# q2 plus FILTER condition
CONSTRUCT {?artist mo:performed ?performance .
           ?track mo:track_number ?num }
WHERE{
	{?performance mo:performer ?artist}
	{?performance mo:recorded_as ?signal}
	{?signal mo:published_as ?track}
	{?track mo:track_number ?num}
	FILTER (?num > 1 && ?num < 5 )}
\end{lstlisting}
\label{qe:q7}
\end{myquery}

\begin{myquery}
Artists and their name. If available, also retrieves images of the artist, biographic information, and other entries that represent the same artist. The location of the artist must be an IRI.
 \lstset{language=SPARQL,framesep=4pt,basicstyle=\scriptsize, showstringspaces=false,
 tabsize=1,numbers=none,numberstyle=\small, stepnumber=1, numbersep=5pt}
\begin{lstlisting}
# q3 plus FILTER condition

CONSTRUCT {?artist foaf:name ?name; 
           foaf:img ?img;
           mo:biography ?bio; 
           bio:olb ?olb;
           owl:sameAs ?artist2;
           foaf:based_near ?p }
WHERE {
         ?artist a mo:MusicArtist ;
         foaf:name ?name .
OPTIONAL { ?artist foaf:img ?img}.
OPTIONAL { ?artist mo:biography ?bio}.
OPTIONAL { ?artist bio:olb ?olb}.
OPTIONAL { ?artist owl:sameAs ?artist2 }.
OPTIONAL { ?artist foaf:based_near ?p .
           FILTER (!isIRI(?p))}
}
\end{lstlisting}
\label{qe:q8}
\end{myquery}

\begin{myquery}
Artist and records, where the artist has made the record and its location is not USA or the record was made by the artist.
 \lstset{language=SPARQL,framesep=4pt,basicstyle=\scriptsize, showstringspaces=false,
 tabsize=1,numbers=none,numberstyle=\small, stepnumber=1, numbersep=5pt}
\begin{lstlisting}
# q4 plus FILTER condition

CONSTRUCT {?artist foaf:made ?record}
WHERE{
	{?artist a mo:MusicArtist .
	?record a mo:Record .
	?record foaf:maker ?artist .
	?artist foaf:based_near ?place .
	FILTER (?place != <http://dbpedia.org/resource/USA>)
	} UNION
	{?artist a mo:MusicArtist .
	?record a mo:Record .
	?artist foaf:made ?record }
}
\end{lstlisting}
\label{qe:q9}
\end{myquery}

\begin{myquery}
Artist and works, where the artist has made that work and this information exists in the Jamendo dataset.
Artist name and works, where the work has been made by the artist, and the artist name begins with ``the'' and this information exists in the Magnatune dataset.
 \lstset{language=SPARQL,framesep=4pt,basicstyle=\scriptsize, showstringspaces=false,
 tabsize=1,numbers=none,numberstyle=\small, stepnumber=1, numbersep=5pt}
\begin{lstlisting}
# q5 plus FILTER condition

CONSTRUCT {?artist1 foaf:made ?work1 .
           ?name2 foaf:made ?work2}
FROM NAMED <http://dbtune.org/jamendo>
FROM NAMED <http://dbtune.org/magnatune>
WHERE
{	                
  GRAPH <http://dbtune.org/jamendo>{
    ?artist1 foaf:made ?work1 } .
  GRAPH <http://dbtune.org/magnatune> {
	?work2 foaf:maker ?artist2 .
    ?artist2 foaf:name ?name2 .
    FILTER (REGEX(str(?name2), ``^the'', ``i''))}
}

\end{lstlisting}
\label{qe:q10}
\end{myquery}

\subsection*{Group C: queries in Group B plus negation} 

\begin{myquery}
Artists and the records they have made, only for artists which name begins with ``the'' and for which no biographical information is stated.
 \lstset{language=SPARQL,framesep=4pt,basicstyle=\scriptsize, showstringspaces=false,
 tabsize=1,numbers=none,numberstyle=\small, stepnumber=1, numbersep=5pt}
\begin{lstlisting}
# q6 plus negation

CONSTRUCT {?artist foaf:made ?record}
WHERE{
	?artist a mo:MusicArtist .
	?record a mo:Record .
	?record foaf:maker ?artist .
	?artist foaf:name ?name .
	FILTER (REGEX(str(?name), ``^the'', ``i'')).
	OPTIONAL {?artist mo:biography ?bio}.
	FILTER (!BOUND(?bio))
}
\end{lstlisting}
\label{qe:q11}
\end{myquery}

\begin{myquery}
Artists and their performances, where the performance has been recorded and published as a track with a track number and the track number is between 1 and 5, but no information can be found regarding the chart position of the track.
 \lstset{language=SPARQL,framesep=4pt,basicstyle=\scriptsize, showstringspaces=false,
 tabsize=1,numbers=none,numberstyle=\small, stepnumber=1, numbersep=5pt}
\begin{lstlisting}
# q7 plus negation
CONSTRUCT {	?artist mo:performed ?performance .
			?track mo:track_number ?num }
WHERE{
	{?performance mo:performer ?artist}
	{?performance mo:recorded_as ?signal}
	{?signal mo:published_as ?track}
	{?track mo:track_number ?num}
	FILTER (?num > 1 && ?num < 5 )
	OPTIONAL {?track mo:chart_position ?pos}.
	FILTER (!BOUND(?pos))	}
\end{lstlisting}
\label{qe:q12}
\end{myquery}

\begin{myquery}
Artists and their name. If available, also retrieves images of the artist, biographic information and location. The location of the artist must be an IRI and no other artist should be stated as the same.
 \lstset{language=SPARQL,framesep=4pt,basicstyle=\scriptsize, showstringspaces=false,
 tabsize=1,numbers=none,numberstyle=\small, stepnumber=1, numbersep=5pt}
\begin{lstlisting}
# q8 plus negation

CONSTRUCT {?artist foaf:name ?name; 
           foaf:img ?img;
           mo:biography ?bio; 
           bio:olb ?olb;
           foaf:based_near ?p }
WHERE {
 ?artist a mo:MusicArtist ;
         foaf:name ?name .
OPTIONAL { ?artist foaf:img ?img}.
OPTIONAL { ?artist mo:biography ?bio}.
OPTIONAL { ?artist bio:olb ?olb}.
OPTIONAL { ?artist owl:sameAs ?artist2 .
           FILTER (!BOUND(?artist2))}.
OPTIONAL { ?artist foaf:based_near ?p . 
           FILTER (!isIRI(?p))}
}
\end{lstlisting}
\label{qe:q13}
\end{myquery}

\begin{myquery}
Artist and records, where the artist has made the record and its location is not USA or the record was made by the artist but it is not available in any kind of support.
 \lstset{language=SPARQL,framesep=4pt,basicstyle=\scriptsize, showstringspaces=false,
 tabsize=1,numbers=none,numberstyle=\small, stepnumber=1, numbersep=5pt}
\begin{lstlisting}
# q9 plus negation

CONSTRUCT {?artist foaf:made ?record}
WHERE{
	{?artist a mo:MusicArtist .
	?record a mo:Record .
	?record foaf:maker ?artist .
	?artist foaf:based_near ?place .
	FILTER (?place != 
	   <http://dbpedia.org/resource/USA>)
	} UNION
	{?artist a mo:MusicArtist .
	?record a mo:Record .
	?artist foaf:made ?record .
	OPTIONAL {?record mo:available_as ?support }.
	FILTER (!BOUND(?support))}
}
\end{lstlisting}
\label{qe:q14}
\end{myquery}

\subsection*{Group D: queries in Group C plus \texttt{ORDER BY} expressions} 

\begin{myquery}
Artists and the records they have made, only for artists whose name begins with ``the'' and for whom no biographical information is stated. The results are sorted by artist.
 \lstset{language=SPARQL,framesep=4pt,basicstyle=\scriptsize, showstringspaces=false,
 tabsize=1,numbers=none,numberstyle=\small, stepnumber=1, numbersep=5pt}
\begin{lstlisting}
# q11 plus ORDER BY

CONSTRUCT {?artist foaf:made ?record}

WHERE{
	?artist a mo:MusicArtist .
	?record a mo:Record .
	?record foaf:maker ?artist .
	?artist foaf:name ?name .
	
	FILTER (REGEX(str(?name), ``^the'', ``i'')).
	
	OPTIONAL {?artist mo:biography ?bio}.
	FILTER (!BOUND(?bio))
}
ORDER BY ?artist


\end{lstlisting}
\label{qe:q15}
\end{myquery}

\begin{myquery}
Artists and their performances, where the performance has been recorded and published as a track with a track number and the track number is between 1 and 5, but no information can be found regarding the chart position of the track.
The results are ordered by artist and track number.
 \lstset{language=SPARQL,framesep=4pt,basicstyle=\scriptsize, showstringspaces=false,
 tabsize=1,numbers=none,numberstyle=\small, stepnumber=1, numbersep=5pt}
\begin{lstlisting}
# q12 plus ORDER BY

CONSTRUCT {	?artist mo:performed ?performance .
			?track mo:track_number ?num }
			
WHERE{
	{?performance mo:performer ?artist}
	{?performance mo:recorded_as ?signal}
	{?signal mo:published_as ?track}
	{?track mo:track_number ?num}
	
	FILTER (?num > 1 && ?num < 5 )
	
	OPTIONAL {?track mo:chart_position ?pos}.
	FILTER (!BOUND(?pos))
}
ORDER BY ?artist ?num


\end{lstlisting}
\label{qe:q16}
\end{myquery}

\begin{myquery}
Artists and their name. If available also retrieves images of the artist, biographic information and location. The location of the artist must be an IRI and no other artist is reported as the same one (i.e., through the owl:sameAs predicate). The results are ordered by artist.
 \lstset{language=SPARQL,framesep=4pt,basicstyle=\scriptsize, showstringspaces=false,
 tabsize=1,numbers=none,numberstyle=\small, stepnumber=1, numbersep=5pt}
\begin{lstlisting}
# q13 plus ORDER BY

CONSTRUCT {?artist foaf:name ?name; 
           foaf:img ?img;
           mo:biography ?bio; 
           bio:olb ?olb;
           owl:sameAs ?artist2;
           foaf:based_near ?p }
WHERE {
 ?artist a mo:MusicArtist ;
         foaf:name ?name .
OPTIONAL { ?artist foaf:img ?img}.
OPTIONAL { ?artist mo:biography ?bio}.
OPTIONAL { ?artist bio:olb ?olb}.
OPTIONAL { ?artist owl:sameAs ?artist2 . 
           FILTER (!BOUND(?artist2))}.
OPTIONAL { ?artist foaf:based_near ?p . 
           FILTER (!isIRI(?p))}
}
ORDER BY DESC(?artist)

\end{lstlisting}
\label{qe:q17}
\end{myquery}

\begin{myquery}
Artist and records, where the artist has made the record and its location is not USA or the record was made by the artist. The results are ordered by artist.
 \lstset{language=SPARQL,framesep=4pt,basicstyle=\scriptsize, showstringspaces=false,
 tabsize=1,numbers=none,numberstyle=\small, stepnumber=1, numbersep=5pt}
\begin{lstlisting}
# q14 plus ORDER BY

CONSTRUCT {?artist foaf:made ?record}
WHERE{
	{?artist a mo:MusicArtist .
	?record a mo:Record .
	?record foaf:maker ?artist .
	?artist foaf:based_near ?place .
	FILTER (?place != 
	     <http://dbpedia.org/resource/USA>)
	} UNION
	{?artist a mo:MusicArtist .
	?record a mo:Record .
	?artist foaf:made ?record }
}
ORDER BY DESC(?artist)

\end{lstlisting}
\label{qe:q18}
\end{myquery}

\section{Schema Information Extraction}
\label{sec:app1}

In this appendix we present the queries performed to extract schema information from the selected datasets. The extracted information was used to produce the graphical representation depicted in Figure~\ref{fig:sourceSch}. First, let define some sets of triples:

\begin{mydef}[Notation]
Let $BT$, $MT$ and $JT$ be the sets of triples from the BBC, Magnatune and Jamendo datasets, respectively. 
Let $MO$ be the set of triples resulting of the extraction of RDFS data from the OWL MusicOntology.
Let $D=BT \cup MT \cup JT $
\end{mydef}

We begin by retrieving all the classes used in $D$. In order to do so we formulate the following SPARQL query, which retrieves all the elements of $B \cup U \cup L$ (as defined in Section~\ref{sec2:rdf}) which appear as object in any triple that uses \texttt{rdf:type} as predicate. Let us call $C$ the resulting collection. For each $c \in C$ we create a light grey node labeled $c$. Light grey nodes represent classes used in the dataset $D$.

\begin{myqueryset}
\label{qe:q1}
 \lstset{language=SPARQL,framesep=4pt,basicstyle=\scriptsize, showstringspaces=false, tabsize=1,numbers=none}
\begin{lstlisting}
SELECT DISTINCT ?c
FROM D
WHERE {?s rdf:type ?c)

\end{lstlisting}
\end{myqueryset}

Let us now retrieve predicates that are used to relate class instances. For this we formulate the following query and store its results in the graph $P1$. For each triple $ (c1,p,c2) \in P1$ we create an arc labeled $p$ from node labeled $c1$ to node labeled $c2$. Directed arcs represent properties used in the dataset $D$.

\begin{myqueryset}
\label{qe:q1}
 \lstset{language=SPARQL,framesep=4pt,basicstyle=\scriptsize, showstringspaces=false, tabsize=1,numbers=none}
\begin{lstlisting}
CONSTRUCT {?c1 ?p ?c2}
FROM D
WHERE {?s1 ?p ?s2 . 
       ?s1 rdf:type ?c1 . 
       ?s2 rdf:type ?c2}

\end{lstlisting}
\end{myqueryset}

We must now retrieve all the sub-classes and super-classes in the $MO$ of classes in $C$. We formulate the following query, storing its results in $C'$. For each $c' \in C'$ we create a dark grey node labeled $c'$. Dark grey nodes represent classes from the MusicOntology 
hierarchically related to classes in $D$.

\begin{myqueryset}
\label{qe:q1}
 \lstset{language=SPARQL,framesep=4pt,basicstyle=\scriptsize, showstringspaces=false, tabsize=1,numbers=none}
\begin{lstlisting}
SELECT DISTINCT ?c1
FROM D
FROM MO
WHERE { 
       { ?s rdf:type ?c . 
         ?c rdfs:subClassOf ?c1} UNION
       { ?s rdf:type ?c . 
         ?c1 rdfs:subClassOf ?c}
}

\end{lstlisting}
\end{myqueryset}

To generate the arcs between classes from the MusicOntology and classes in $D$ we formulate the following query, storing its results in graph $P2$. For each triple $ (c1,rdfs:subClassOf,c2) \in P2$ we create a dashed arc from node labeled $c1$ to node labeled $c2$. Dashed arcs represent rdfs:subClassOf properties.

\begin{myqueryset}
\label{qe:q1}
 \lstset{language=SPARQL,framesep=4pt,basicstyle=\scriptsize, showstringspaces=false, tabsize=1,numbers=none}
\begin{lstlisting}
CONSTRUCT {?c rdfs:subClassOf ?c1}
FROM D
FROM MO
WHERE { 
       { ?s rdf:type ?c . 
         ?c rdfs:subClassOf ?c1} UNION
       { ?s rdf:type ?c1 . 
         ?c rdfs:subClassOf ?c1}
}

\end{lstlisting}
\end{myqueryset}

Finally we want to retrieve used predicates that have literals as objects. To do so we formulate the following query, storing its results in $P3$. For each pair $(p,c) \in P3$ we create a label $p$ next to node $c$. Labels next to nodes represents properties whose
range is not a class.

\begin{myqueryset}
\label{qe:q1}
 \lstset{language=SPARQL,framesep=4pt,basicstyle=\scriptsize, showstringspaces=false, tabsize=1,numbers=none}
\begin{lstlisting}
SELECT DISTINCT ?p ?c
FROM D
WHERE  {?s1 ?p ?s2 . 
        ?s1 rdf:type ?c . 
        {OPTIONAL {?s2 rdf:type ?a2} . 
                   NOT BOUND(?a2)}
}

\end{lstlisting}
\end{myqueryset}

\onecolumn
\section{Datasets Selection}
\label{sec:app2}

In this appendix we provide insight about the selection process of datasets.
Table~\ref{tab:dataSelection} presents detailed information on the available datasets.

\begin{table*}[ht!]
 \caption{Description of available datasets at LOD site}
  \label{tab:dataSelection}
\begin{tabular}{| p{0.5cm} | p{4cm} | p{7cm} | p{3cm}|}
\hline
\textbf{Nr } &\textbf{Project name } & \textbf{Data domain} & \textbf{Size of Data Set }  \\ \hline
1 &  Allen Brain Atlas  & Brain data & 51 MB  \\ \hline
2 &  Airport Data  & Airport data & \textgreater 750 k triples \\ \hline
3 &  BAMS  & Brain data & 5.6 MB  \\ \hline
4 &  BBC John Peel sess  & Music data & \textgreater 270 k triples  \\ \hline
5 &  BBOP  & Various bio- and gene- related datasets  & 36 MB  \\ \hline
6 &  BTC Datasets  & Various & \textgreater 2 billion triples  \\ \hline
7 &  Bio2RDF  & Various bio- and gene- related datasets  & 2.7 billion triples  \\ \hline
8 &  Bitzi  & Digital media data & \textgreater 300 K files, 270MB uncompressed  \\ \hline
9 &  Data-gov Wiki  & Gubernamental data  & \textgreater 5 billion triples  \\ \hline
10 &  DBpedia  & Various data extracted from Wikipedia & 247 million triples  \\ \hline
11 &  Entrez Gene  & Gene data & 7.7 MB  \\ \hline
12 &  Freebase  & Various data extracted from Freebase  & 505 MB compressed \\ \hline
13 &  GeoSpecies KB  & Information on Biological Orders, Families, Species & 1.888 M triples  \\ \hline
14 &  GO annotations  & Gene data & 73 MB  \\ \hline
15 &  GovTrack.us  & Data about the U.S. congress  & 13 million triples  \\ \hline
16 &  Jamendo  & Music data  & 1.1 million triples  \\ \hline
17 &  LinkedCT  & Clinical traits data & 9.8 million triples, 1.6GB  \\ \hline
18 &  LinkedMDB  & Linked Data about Movies  & 6.1 million triples, 850MB \\ \hline
19 &  Linked Sensor Data  & Weather sensor data & 1.7 billion triples  \\ \hline
20 &  Magnatune  & Music data & \textgreater 400 k triples, 40 MB \\ \hline
21 &  MeSH headings  & Medline papers data & 758 MB  \\ \hline
22 &  MusicBrainz  & Music data & N/A \\ \hline
23 &  OpenCyc  & OpenCyc Ontology  & \textgreater 1.6 million triples, \textgreater 150MB uncompressed  \\ \hline
24 &  RKB Explorer Data  & 25 different domains, each with a separate data set. Scientific research & \textgreater 60 million triples  \\ \hline
25 &  STW Thesaurus for Economics  & Thesaurus for economics and business economics & 12 MB uncompressed  \\ \hline
26 &  SwetoDblp  & Ontology focused on bibliography data of publications from DBL & 11M triples  \\ \hline
27 &  TaxonConcept KB  & Species Concepts and related Biodiversity Informatics data  & 8.2M triples  \\ \hline
28 &  Telegraphis LOD  & Geographic data from GeoNames and Wikipedia data  & \textless 10k triples a piece  \\ \hline
29 &  TCMGeneDIT  & Traditional Chinese medicine, gene and disease association dataset and a linkset mapping TCM gene symbols to Extrez Gene IDs created by Neurocommons  & 288kb compressed  \\ \hline
30 &  t4gm.info  & Thesaurus for Graphic Materials  & 7.3MB uncompressed  \\ \hline
31 &  UniProt  & a large life sciences data set  & \textgreater 300M triples  \\ \hline
32 &  U.S. Census  & population statistics from the U.S & 1 billion triples  \\ \hline
33 &  U.S. SEC  & corporate ownership  & 1.8 million triples  \\ \hline
34 &  YAGO  & Data from different sources (Wikipedia, WordNet, GeoNames) focused on persons, organizations, etc. & 1Gb  \\ \hline
\end{tabular}
\end{table*}

Table~\ref{tab:reqEval} presents the results of the evaluation of the requirements stated in Section ~\ref{sec5:datasets} for each dataset in Table ~\ref{tab:dataSelection}. Information regarding requirement 5 is only stated if available or if the other requirements are fulfilled, otherwise it is stated as N/A (not available).

\begin{table*}[htbp]
 \caption{Requirement evaluation for each dataset}
  \label{tab:reqEval}
\begin{tabular}{|l|l|l|l|l|l|l|}
\hline
\textbf{Nr} &\textbf{Dataset} & \textbf{req1:domain} & \textbf{req2:heterog} & \textbf{req3:size} & \textbf{req4:dump} & \textbf{req5:RDFS} \\ \hline
1 & Allen Brain Atlas  & no & no & no & yes & N/A \\ \hline
2 & Airport Data  & yes & no & yes & no & N/A \\ \hline
3 & BAMS  & no & no & no & yes & N/A \\ \hline
4 & BBC John Peel sess & yes & yes & yes & yes & OWL \\ \hline
5 & BBOP  & no & yes & no & yes & N/A \\ \hline
6 & BTC Datasets & +/- & yes & yes & yes & yes \\ \hline
7 & Bio2RDF  & no & yes & yes & yes & N/A \\ \hline
8 & Bitzi  & yes & no & yes & no & N/A \\ \hline
9 & Data-gov Wiki  & yes & yes & yes & yes & N/A \\ \hline
10 & DBpedia  & yes & no & yes & yes & no \\ \hline
11 & Entrez Gene  & no & yes & no & no & N/A \\ \hline
12 & Freebase  & yes & no & yes & no & N/A \\ \hline
13 & GeoSpecies KB & no & no & yes & yes & N/A \\ \hline
14 & GO annotations & no & yes & no & no & N/A \\ \hline
15 & GovTrack.us  & yes & yes & yes & yes & N/A \\ \hline
16 & Jamendo & yes & yes & yes & yes & OWL \\ \hline
17 & LinkedCT  & no & yes & yes & yes & N/A \\ \hline
18 & LinkedMDB  & yes & no & yes & yes & no \\ \hline
19 & Linked Sensor Data  & yes & no & yes & yes & OWL \\ \hline
20 & Magnatune & yes & yes & yes & yes & OWL \\ \hline
21 & MeSH headings  & no & yes & yes & yes & N/A \\ \hline
22 & MusicBrainz  & yes & yes & N/A & no & N/A \\ \hline
23 & OpenCyc  & no & no & yes & no & N/A \\ \hline
24 & RKB Explorer Data  & yes & yes & yes & no & N/A \\ \hline
25 & STW Thesaurus for Economics  & no & no & no & yes & N/A \\ \hline
26 & SwetoDblp  & yes & no & no & yes & OWL \\ \hline
27 & TaxonConcept KB & no & no & no & yes & N/A \\ \hline
28 & Telegraphis LOD  & yes & yes & no & yes & N/A \\ \hline
29 & TCMGeneDIT  & no & no & no & yes & N/A \\ \hline
30 & t4gm.info  & yes & no & no & yes & N/A \\ \hline
31 & UniProt  & no & yes & yes & yes & N/A \\ \hline
32 & U.S. Census  & yes & no & yes & no & N/A \\ \hline
33 & U.S. SEC  & yes & no & yes & yes & N/A \\ \hline
34 & YAGO  & yes & no & yes & yes & RDFS \\ \hline
\end{tabular}
\label{}
\end{table*}


\end{document}